\documentclass[tightenlines,twoside,secnumarabic,
               onecolumn,floatfix,nofootinbib,showpacs,11pt]{revtex4}
\usepackage[titletoc,title]{appendix}
\usepackage{comment}
\usepackage[english]{babel}
\usepackage{amsmath,amssymb}
\usepackage{graphicx}
\usepackage[sort&compress]{natbib}
\usepackage{slashed}            
\allowdisplaybreaks

\bibpunct{[}{]}{,}{n}{}{,}
\def\spa#1.#2{\left\langle#1#2\right\rangle}
\def\spb#1.#2{\left[#1#2\right]}
\def\spab#1.#2.#3{\left\langle#1|#2|#3\right]}
\def\spba#1.#2.#3{\left[#1|#2|#3\right\rangle}
\def\spaa#1.#2.#3.#4{\left\langle#1|#2|#3|#4\right\rangle}
\def\spbb#1.#2.#3.#4{\left[#1|#2|#3|#4\right]}
\def\lsim{\mathrel{\rlap{\lower4pt\hbox{\hskip1pt$\sim$}}
    \raise1pt\hbox{$<$}}}                
\def\gsim{\mathrel{\rlap{\lower4pt\hbox{\hskip1pt$\sim$}}
    \raise1pt\hbox{$>$}}}                

\usepackage{comment}
\usepackage{color}
\newcommand{\be}{\begin{equation}}
\newcommand{\ee}{\end{equation}}
\newcommand{\fb}{\,{\rm fb}}

\newcommand{\gev}{{\rm GeV}}

\def\Li{\mathrm{Li}}

\def\bea{\begin{eqnarray}}
\def\eea{\end{eqnarray}}
\def\ltap{\ \raise.3ex\hbox{$<$\kern-.75em\lower1ex\hbox{$\sim$}}\ }
\def\gtap{\ \raise.3ex\hbox{$>$\kern-.75em\lower1ex\hbox{$\sim$}}\ }
\def\lsim{\ \raise.3ex\hbox{$<$\kern-.75em\lower1ex\hbox{$\sim$}}\ }
\def\gsim{\ \raise.3ex\hbox{$>$\kern-.75em\lower1ex\hbox{$\sim$}}\ }
\def\eg{{\it e.g.}}
\def\ie{{\it i.e.}}
\def\etc{{\it etc}}
\usepackage{slashed}
\newcommand{\met}{\slashed {E}_{T}}
\usepackage{color}

\usepackage{xcolor}
\definecolor{red}{rgb}{1.0, 0, 0}

\begin{document}

\title{Next-to-Leading Order Predictions for Dark Matter Production at Hadron Colliders}
\author{Patrick J. Fox }          \email[Email: ]{pjfox@fnal.gov}
\author{Ciaran Williams}            \email[Email: ]{ciaran@fnal.gov}
\affiliation{Theoretical Physics Department \\
                  \mbox{Fermilab, P.O.~Box 500, Batavia, IL 60510, USA} \vspace{0.2cm}\\
 }
\date{\today} 
\pacs{}

\begin{abstract}
We provide Next-to-Leading Order (NLO) predictions for Dark Matter (DM) production in association with either a jet
or a photon at hadron colliders. In particular we study the production of a pair of fermionic DM particles through a mediator which couples to SM via either a vector, axial-vector, scalar, pseudo-scalar, or gluon-induced coupling. Experimental constraints on the scale of new physics associated with these operators are limited by systematics, highlighting the need for NLO signal modeling. 
We factorize the NLO QCD and the DM parts of the calculation, allowing the possibility of using the results presented here for a large variety of searches in monojet and monophoton final states. Our results are implemented into the Monte Carlo program MCFM. 
\end{abstract}
\pacs{}
\date{\today}

\begin{flushright}
  FERMILAB-PUB-12-612-T
\end{flushright}

\maketitle
\newpage

\section{Introduction} 

While the existence of Dark Matter (DM) has now been well established, it is still unknown what object, or objects, makes up this dominant component of the matter in the universe.  Perhaps the best motivated candidate is a new neutral particle whose mass is around the weak scale ($\mathcal{O}(1-1000)\ \gev$) and whose couplings to Standard Model (SM) fields are somewhat below the weak scale.  So far direct detection experiments have seen no clear signal consistent with DM recoiling against atoms~\cite{Aprile:2011hi,Ahmed:2009zw,Behnke:2012ys}, with several notable exceptions~\cite{Bernabei:2010mq,Aalseth:2011wp,Angloher:2011uu}.  At the same time the LHC is performing a multitude of searches, many of which require substantial missing energy.  Although these searches are sensitive to the production of DM, most are designed to search for models where the DM is made in the cascade decay of a colored particle, \eg ~supersymmetric searches, and are not generally applicable to all models. 

Recently, it has been noted that there is a class of more model independent searches for DM production that can be carried out at the LHC\footnote{For constraints on DM coming from other colliders see \cite{Bai:2010hh,Fox:2011fx}.}, based around the production of DM in association with monojets \cite{Beltran:2010ww,Goodman:2010yf,Goodman:2010ku,Rajaraman:2011wf,Fox:2011pm,Shoemaker:2011vi,Fortin:2011hv}, monophotons \cite{Fox:2011pm,Fortin:2011hv}, mono-$W$~\cite{Bai:2012xg} and also a more inclusive multijet search~\cite{Fox:2012ee}.  This class of searches can place strong constraints on the properties of DM that are complementary to those from direct detection searches.  In particular, the LHC does not suffer from a low mass threshold, nor is the spin-dependent (SD) bound considerably weaker than the spin-independent (SI).  Although the collider SI bounds for DM mass above $\sim 10\ \gev$ are weaker than direct detection, the SD bounds are often stronger.  In addition, the DM is being produced by the experiment rather than relying on a galactic component in our vicinity and so is not sensitive to unknown astrophysics.

A dedicated DM search, using the shape of the leading jet transverse momentum, $p_T$, distribution has been carried out at CDF on 6.7 fb$^{-1}$ of data~\cite{Aaltonen:2012jb}.  CMS and ATLAS have both carried out cut-and-count based monojet \cite{Chatrchyan:2012me,ATLAS:2012ky} and monophoton \cite{Chatrchyan:2012tea,Aad:2012fw} analyses using $\sim 5$ fb$^{-1}$ of data.  Given the copious rate for the production of $Z$'s in association with jets the current round of experimental results are already dominated by systematic (rather than statistical) forms of uncertainty. In the near future this will also be the case for the smaller (but by no means small) production rate of $Z$ in association with a photon.  Given that the analyses are therefore systematics limited the theoretical community should investigate the possibilities in which the systematic errors associated with the theoretical predictions can be reduced. 

The most obvious mechanism to reduce the theoretical systematic uncertainty is to provide Next-to-Leading Order (NLO) predictions for the irreducible (and reducible) backgrounds. However, NLO predictions for $Z + (j/\gamma)$ have been available for over a decade~\cite{Giele:1991vf,Baur:1997kz}. In addition, recent developments in matching showers to NLO predictions have resulted in a publicly released matched shower prediction for the $Zj$ process within the POWHEG-BOX formalism~\cite{Alioli:2010qp}. Therefore, until the completion of the NNLO $Z + (j/\gamma)$ cross sections potential improvements in the theoretical predictions for the dominant backgrounds are limited (although very recently, EW corrections to monojet production have been computed~\cite{Denner:2012ts}). 

The remaining scope for reducing the theoretical systematic errors resides in improvements in the modeling of the signal process. Thus far experimental analyses have relied on Leading Order (LO) predictions for the signal and as a result are exposed to the inherent issues associated with a LO prediction, namely a large uncertainty in rate and shape (the shape issue can be improved if matched shower predictions are used). By improving the theoretical predictions for DM production by including NLO corrections one therefore reduces the overall scale dependence (reducing the rate uncertainty) and typically obtains a larger value for the cross section. Therefore, a limit obtained using a NLO prediction for the total rate is both stronger (since the cross sections are larger) and more accurate (since the rate uncertainty is reduced).

Furthermore, the typical monojet search cuts on the missing transverse momentum, $\met$, and the leading jet $p_T$ are usually mismatched
and the searches employed at the LHC are more inclusive than the monojet names suggests, allowing up to two jets in the event. Thus, there is a region of phase space that has two jets whose individual $p_T$ is below the $\met$ cut, but above the jet $p_T$ requirement, which together recoil against the DM, that is not included at LO.  This contribution is naturally included in predictions which have access to the DM plus two-jet phase space, \eg~NLO and matched shower predictions.

For these reasons we therefore implement the production of a pair (of fermionic) DM particles in association with either at jet or a photon into the MCFM~\cite{MCFMweb,Campbell:1999ah,Campbell:2011bn} code, which is available publicly. We will study a range of phenomenologically interesting operators, primarily (but not restricted to) the effective theory in which the particle responsible for mediating the DM/SM interaction is very heavy. We will present our results in a format which allows easy recycling of our results, more specifically we present the results for the SM and DM parts of the calculation in a factorized manner. This allows easy implementation of our results to other new physics scenarios which produce monojet or monophoton signatures. 

This paper proceeds as follows, in section~\ref{sec:DMcalc} we provide the necessary DM and SM ingredients to construct our NLO calculation. In section~\ref{sec:monojet} we investigate the NLO phenomenology associated with the monojet final state. Section~\ref{sec:monogamma} presents a similar study for the monophoton channel. We draw our conclusions in section~\ref{sec:conclusions}. In appendix~\ref{app:spin} we include some spinor definitions and appendix~\ref{app:VecC} catalogues some of the formulae obtained during our calculation.

\section{NLO Calculations of DM production} 
\label{sec:DMcalc} 

The strategy we will follow to provide NLO predictions for monophoton and monojet DM processes is to 
factorize the problem into SM production and followed by a BSM ``decay'' in the final state.  That is, we will consider operators of the form $\mathcal{O}_{SM}\mathcal{O}_{DM}$ where the particle coupling the SM to the dark sector is exchanged in the $s$-channel. For instance, the amplitude for monojet production through the vector operator is of the form,
\be
A_V(1_q,2_g,3_{\bar{q}},4_{\bar{\chi}},5_\chi) = \mathcal{A}_\mu(1_q,2_g,3_{\bar{q}})\times \mathcal{V}^\mu(4_{\bar{\chi}},5_\chi)~.
\ee 
In the following subsections, and Appendix~\ref{app:VecC}, we give expressions for both $\mathcal{A}$ and $\mathcal{V}$.

Situations involving $t$-channel mediators, for instance squark exchange, can still be considered by carrying out a Fierz transformation. These will then involve contributions from multiple $s$-channel operators.  Our implementation in MCFM allows for this full generality but we do not consider such a case here.  We will be predominantly interested in DM production which proceeds via an effective field theory (EFT), where the exchanged particle is integrated out. We we also briefly provide examples for the case of a light mediator for a subset of our operators. The effective theory is well motivated provided that the mediating particles are heavy ($\gtrsim $ few TeV), however whether or not the mediating propagator is included bears little impact on the NLO calculations, our MCFM implementation can calculate either in the EFT or the full theory.  The operators we consider that involve SM quarks correspond to vector, axial-vector, scalar and pseudo-scalar exchange, and we consider one operator that couples gluons to DM,
\begin{eqnarray}
\mathcal{O}_V&=&\frac{(\overline{\chi}\gamma_{\mu}\chi)(\overline{q}\gamma^{\mu}q)}{\Lambda^2}~,\label{eq:OV}  \\
\mathcal{O}_A&=&\frac{(\overline{\chi}\gamma_{\mu}\gamma_5\chi)(\overline{q}\gamma^{\mu}\gamma_5q)}{\Lambda^2}~,\label{eq:OA}\\
\mathcal{O}_g&=&\alpha_s\frac{(\chi\overline{\chi})(G^{\mu\nu}_aG_{a,\mu\nu})}{\Lambda^3}~,\label{eq:Og} \\
\mathcal{O}_S&=&\frac{m_q(\overline{\chi}\chi)(\overline{q}q)}{\Lambda^3}~,\label{eq:OS}  \\
\mathcal{O}_{PS}&=&\frac{m_q(\overline{\chi}\gamma_5\chi)(\overline{q}\gamma_5q)}{\Lambda^3}\label{eq:OPS}~.
\end{eqnarray}
We restrict our focus to the above operators, which provide a representative sample of the phenomenologically interesting models.  Due to our factorization approach, more general cases, say for instance an operator of the form $(\overline{\chi}\gamma_{\mu}\chi)(\overline{q}\gamma^{\mu}\gamma_5q)$, can be readily obtained from the results we will present. The operator $\mathcal{O}_{V(A)}$ has a simple UV completion involving exchange of a vector (axial-vector) boson, of mass $M$ and width $\Gamma$.  The full theory corresponds to the replacement $\Lambda^2\rightarrow (s_{\bar{\chi}\chi}-M^2+i M\Gamma)/g_\chi g_q$.  Operator $\mathcal{O}_g$ is induced at the loop level and the simplest UV completion involves a heavy scalar and heavy fermions.  The scale of the new physics, $M$, is typically lower than for $\mathcal{O}_{V(A)}$ and so the effective theory has a smaller range of validity.

For the scalar and pseudo-scalar operators we have written the couplings as scaling with quark mass.  This is what is expected if an assumption of minimal flavour violation (MFV) is made, in which case the only flavour violating spurions are the Yukawa matrices.  With $SU(2)$ invariance requiring an implicit Higgs field insertion the operators then scale with quark mass.  Here we only consider the flavour diagonal part of these operators, flavour violation in DM couplings leads to other interesting signals~\cite{Andrea:2011ws,Kamenik:2011nb}.  If the MFV assumption is loosened one has to contend with strong constraints from flavour observables.  One possibility would be if the DM couplings were $\mathcal{O}(1)$ in the quark mass eigenstate basis, although such a model would be highly tuned.  Our MCFM implementation is sufficiently flexible to allow either possibility.  We focus in this paper on the more motivated MFV case and present results for that.  Because of PDFs and the suppressed light quark couplings, there is a difference between the behaviour for the case of DM coupling to the first five generations of quarks and to the top quark~\cite{Haisch:2012kf}.  Therefore we will address the light quarks ($u,\ldots ,b$) and top separately.  

Finally, unlike in the effective theory, $t$-channel operators present an additional problem upon UV completion.  Since the exchanged state has to be coloured there are additional diagrams that must be included in order to achieve NLO accuracy.  Therefore, for $t$-channel operators our NLO results can only be used for the case where the mediator has been integrated out. 

In order to facilitate the calculation we will simplify the problem further by calculating helicity amplitudes. For the massless SM production amplitudes this introduces a dramatic simplification in the number of independent calculations, since helicity amplitudes naturally involve projections of the form $(1\pm\gamma^5)$. This allows us to determine the results for the various operators from common building blocks. Since the DM particles are massive care must be taken because the mass spoils the chiral symmetry.

\subsection{Dark Currents} 

We wish to use helicity methods in order to calculate our dark currents.  However, since 
the fermionic DM is massive helicity is not a good quantum number. Nevertheless, massive fermions must 
satisfy the usual sum rule when summing over polarization states, 
\begin{eqnarray} 
\sum_{s=\pm} u_s(p,m)\overline{u}_s(p,m) &=& \slashed{p}+m  \\
\sum_{s=\pm} v_s(p,m)\overline{v}_s(p,m) &=& \slashed{p}-m~. 
\end{eqnarray}
Therefore, one can re-write a massive spinor in terms of two massless spinors \cite{Kleiss:1985yh} provided that the completeness relation above is preserved. In order to write massive spinor bilinears in terms of massless ones we use the prescription of refs.~\cite{Rodrigo:2005eu,Badger:2010mg} rewriting momenta ($p_1$ and $p_2$ with $p_i^2=m^2$) for two massive particles, of mass $m$, in terms of two massless particles ($k_1$ and $k_2$, with $k_i^2=0$) such that, 
\begin{eqnarray}
p_1^{\mu}=\frac{1+\beta}{2} k_1^{\mu}+\frac{1-\beta}{2} k_2^{\mu} \\
p_2^{\mu}=\frac{1+\beta}{2} k_2^{\mu}+\frac{1-\beta}{2} k_1^{\mu} 
\end{eqnarray}
where, $\beta=\sqrt{1-4m^2/s_{12}}$, and $s_{12}=(p_1+p_2)^2=2k_1\cdot k_2$.  This decomposition has the advantage that $p_1+p_2=k_1+k_2$.  The helicity states for the massive spinors ($u$, $v$,  \etc) are defined in terms of the massless spinors $|1^\pm\rangle,|2^\pm\rangle$, through the following relations,
\begin{eqnarray}
\overline{u}_{\pm}(p_2,m) = \frac{\beta_+^{-1/2}}{\langle 1^{\mp} | 2^{\pm} \rangle} \langle 1^{\mp}|(\slashed{p}_2+m), \quad {v}_{\pm}(p_1,m) = \frac{\beta_+^{-1/2}}{\langle 1^{\mp} | 2^{\pm} \rangle} (\slashed{p}_1-m)|2^{\pm}\rangle~,
\end{eqnarray}
where we have introduced variables $\beta_{\pm}=\frac{1}{2}(1\pm\beta)$.  A list of spinor definitions is provided in appendix~\ref{app:spin}, and we encourage the interested reader to inspect ref.~\cite{Badger:2010mg} for additional properties and relations of massive spinors. We are now in a position to define the currents we will need for our study, namely those involving the insertion of a $\gamma$ matrix,
\begin{eqnarray}
\mathcal{V}^{\mu}(2^{\pm}_\chi,1^{\mp}_{\overline{\chi}}) &=& \overline{u}_{\pm}(p_2)\gamma^{\mu}v_{\mp}(p_1)  = \langle 2^\pm |\gamma^{\mu} |1^{\mp}\rangle \\
\mathcal{V}^{\mu}(2^{+}_\chi,1^{+}_{\overline{\chi}}) &=& \overline{u}_{+}(p_2)\gamma^{\mu}v_{+}(p_1)  = 2\frac{m}{\spa 1.2} (k_1-k_2)^{\mu}~,\\
\mathcal{V}^{\mu}(2^{-}_\chi,1^{-}_{\overline{\chi}}) &=& \overline{u}_{-}(p_2)\gamma^{\mu}v_{-}(p_1)  =  2\frac{m}{\spb 1.2} (k_1-k_2)^{\mu}~.
\end{eqnarray}
Note that the helicity conserving currents are identical to their massless counterparts, and that the helicity violating currents vanish in the massless limit as required.
For the cases in which the dark matter is mediated through the exchange of an axial vector we will need the following currents, 
\begin{eqnarray}
\mathcal{V}^{\mu}_5(2^{\pm}_\chi,1^{\mp}_{\overline{\chi}}) &=& \overline{u}_{\pm}(p_2)\gamma^{\mu}\gamma_5v_{\mp}(p_1)  = \mp(1-2\beta_+)\langle 2^\pm |\gamma^{\mu} |1^{\mp}\rangle~,  \\
\mathcal{V}^{\mu}_5(2^{+}_\chi,1^{+}_{\overline{\chi}}) &=& \overline{u}_{+}(p_2)\gamma^{\mu}\gamma_5v_{+}(p_1)  = -2\frac{m}{\spa 1.2} (k_1+k_2)^{\mu}~,\\
\mathcal{V}^{\mu}_5(2^{-}_\chi,1^{-}_{\overline{\chi}}) &=& \overline{u}_{-}(p_2)\gamma^{\mu}\gamma_5v_{-}(p_1)  =  +2\frac{m}{\spb 1.2} (k_1+k_2)^{\mu}~.
\end{eqnarray}
In addition to the currents listed above we will also need the following scalar currents, 
\begin{eqnarray}
\mathcal{S}(2^{\pm}_\chi,1^{\mp}_{\overline{\chi}}) &=& \overline{u}_{\pm}(p_2)v_{\mp}(p_1)  = 0~,\\
\mathcal{S}(2^{+}_\chi,1^{+}_{\overline{\chi}}) &=& \overline{u}_{+}(p_2)v_{+}(p_1)  = (2\beta_+ -1)\spb2.1~, \\
\mathcal{S}(2^{-}_\chi,1^{-}_{\overline{\chi}}) &=& \overline{u}_{-}(p_2)v_{-}(p_1)  =   (2\beta_+ -1)\spa 2.1~. 
\end{eqnarray}
Finally we will consider the decays of dark matter through a pseudo-scalar current 
\begin{eqnarray}
\mathcal{S}_5(2^{\pm}_\chi,1^{\mp}_{\overline{\chi}}) &=& \overline{u}_{\pm}(p_2)\gamma_5v_{\mp}(p_1)  = 0~, \\
\mathcal{S}_5(2^{+}_\chi,1^{+}_{\overline{\chi}}) &=& \overline{u}_{+}(p_2)\gamma_5v_{+}(p_1)  = -\spb2.1 ~, \\
\mathcal{S}_5(2^{-}_\chi,1^{-}_{\overline{\chi}}) &=& \overline{u}_{-}(p_2)\gamma_5v_{-}(p_1)  =   \spa 2.1~. 
\end{eqnarray} 
Note that in all cases, the $m\rightarrow 0 $ limit is clearly reproduced correctly. 

\subsection{SM Vector Currents} 

We now define the currents needed to describe the SM production of a gluon and a vector current, $\mathcal{O}_V$. 
Such results are already present in the literature~\cite{Bern:1997sc,Giele:1991vf}, namely amplitudes involving the production of 
a $Z$ (or more precisely a virtual photon $\gamma^*$) and jets. In order to efficiently recycle these amplitudes one must first remove the 
unwanted decay of the $Z$ into two massless leptons. The general strategy is thus to re-write the amplitude for the $Z$ in the form,
\begin{eqnarray}
A^{(0)}_{Zj} (1^+_q,2^+_g,3^-_{\overline{q}},4^{+}_{\ell},5^{-}_{\overline{\ell}})=A^{(0,{\mu})}_{V}(1^+_q,2^+_g,3^-_{\overline{q}})\times J_{\mu}^{Z}(4^{+}_{\ell},5^{-}_{\overline{\ell}}).
\end{eqnarray} 
The current $J_{\mu}^{Z}$ can then be replaced by the DM current of choice to produce the desired result. Similar techniques were used in ref.~\cite{Campbell:2012ft} for the calculation of the amplitudes for $Z\gamma j$ in which the photon was radiated from the final state leptons. 

As an example we present the amplitudes for the tree-level production of a jet (or photon) in association with a vector operator, with momentum $P$ flowing in the $\chi\bar{\chi}$ system, 
\begin{eqnarray}
\mathcal{A}^{(0,\mu)}_{V}(1^+_q,2^+_g,3^-_{\overline{q}})& =& \frac{1}{2}\frac{\spaa 3.(1+2).\gamma^{\mu}.3}{\spa1.2\spa2.3} \\
\mathcal{A}^{(0,\mu)}_{V}(1^+_q,2^-_g,3^-_{\overline{q}})  &=&\mathcal{A}_{V}^{(0,\mu)}(1^+_q,2^+_g,3^-_{\overline{q}})  \,\, ( \spa a.b \leftrightarrow \spb a.b \ , \, (1 \leftrightarrow 3) )~.
\end{eqnarray}
Since the extraction of the currents needed for the NLO corrections are now determined in terms of the known 
results in the literature~\cite{Bern:1997sc,Giele:1991vf} we refrain from writing them explicitly here. However, since they 
are useful building blocks for other monojet searches we present a full list of the currents (for virtual and real corrections) in 
Appendix \ref{app:VecC}. Before moving on to discuss other SM currents we note that the monophoton amplitudes are naturally related to the pieces of the monojet amplitudes which are subleading in colour.

\subsection{SM Scalar Currents} 

In this section we provide the helicity amplitudes for the ``SM'' production of a scalar in association with a jet or a photon. In the previous section we described the extraction of 
the vector currents from the existing literature results for $Z/\gamma^* +$ jet.  However 
this calculation has no obvious analog in the SM since the Higgs couples to massive fermions and we consider five massless flavours. Therefore, we perform the calculation directly.  The NLO calculation of DM plus monophoton for the scalar operator has been studied previously, using traditional matrix element techniques, in \cite{Wang:2011sx}.  We will focus on the case of scalar and pseudo-scalar couplings which are proportional to mass, similar to the SM Higgs couplings and therefore we could have used the $H+b$ results 
from the literature. However we wish to remain general enough to allow for couplings independent of mass, therefore we drop the mass terms in the SM production which are not part of the Yukawa coupling. Since (to the best of our knowledge) the helicity amplitudes for these processes have not been written down before, we present them in full in this section. 

For completeness we note that the case of scalar coupling to the top quark is particularly interesting since there is no tree-level monojet diagram and the LO result is due to a loop of top quarks~\cite{Haisch:2012kf}, these results can be easily extracted from the Higgs plus jet results in MCFM~\cite{EHSB}.

We begin by listing the two independent tree-level amplitudes, 
\begin{eqnarray}
A^{(0)}_S(1^-_q,2^-_{\overline{q}},3^+_g) = \frac{\spa1.2^2}{\spa1.3\spa2.3}~,\\
A^{(0)}_S(1^-_q,2^-_{\overline{q}},3^-_g) = \frac{s_{123}}{\spb1.3\spb2.3}~. 
\end{eqnarray}
The virtual corrections to this amplitude consist of leading and sub-leading colour contributions. The analytic 
forms of these expressions are rather simple, primarily because the insertion of the scalar operator does not 
increase the tensor rank of any loop diagrams. We calculate the four-dimensional cut-constructible pieces using 
the quadruple cut technique~\cite{Britto:2004nc}. We have checked our virtual results 
against a numerical implementation of $D$-dimensional unitarity~\cite{Ellis:2008ir}, finding agreement. This check also 
confirms the lack of bubble pieces from the expressions. The expressions for the virtual amplitudes are very simple, the leading colour pieces have the following form,  
 \begin{eqnarray}
\mathcal{A}^{(1,lc)}_S(1^-_q,2^-_{\overline{q}},3^+_g) &=&c_{\Gamma} \mathcal{A}^{(0)}_S\bigg( -\frac{1}{\epsilon^2}\bigg(\bigg(\frac{\mu^2}{-s_{12}}\bigg)^{\epsilon} +\bigg(\frac{\mu^2}{-s_{23}}\bigg)^{\epsilon}\bigg)+ {\rm{Ls}}_{-1}\bigg(\frac{-s_{13}}{-s_{123}},\frac{-s_{23}}{-s_{123}}\bigg)\bigg)~,\label{eq:s_virtlc}\\
\mathcal{A}^{(1,lc)}_S(1^-_q,2^-_{\overline{q}},3^-_g) &=&c_{\Gamma} \mathcal{A}^{(0)}_S\bigg( -\frac{1}{\epsilon^2}\bigg(\bigg(\frac{\mu^2}{-s_{12}}\bigg)^{\epsilon} +\bigg(\frac{\mu^2}{-s_{23}}\bigg)^{\epsilon}\bigg)+ {\rm{Ls}}_{-1}\bigg(\frac{-s_{13}}{-s_{123}},\frac{-s_{23}}{-s_{123}}\bigg)\bigg)\nonumber\\&&+c_{\Gamma}\frac{s_{13}+s_{23}}{2\spb2.3\spb1.3}~.
\end{eqnarray}
The subleading in colour amplitudes have the following form (these are also the monophoton amplitudes) 
 \begin{eqnarray}
\mathcal{A}^{(1,slc)}_S(1^-_q,2^-_{\overline{q}},3^+_g) &=&c_{\Gamma} \mathcal{A}^{(0)}_S\bigg( -\frac{1}{\epsilon^2}\bigg(\frac{\mu^2}{-s_{12}}\bigg)^{\epsilon}+ {\rm{Ls}}_{-1}\bigg(\frac{-s_{13}}{-s_{123}},\frac{-s_{12}}{-s_{123}}\bigg) \nonumber\\&&+{\rm{Ls}}_{-1}\bigg(\frac{-s_{23}}{-s_{123}},\frac{-s_{23}}{-s_{123}}\bigg)\bigg)~, \\
\mathcal{A}^{(1,slc)}_S(1^-_q,2^-_{\overline{q}},3^-_g) &=&c_{\Gamma} \mathcal{A}^{(0)}_S\bigg( -\frac{1}{\epsilon^2}\bigg(\frac{\mu^2}{-s_{12}}\bigg)^{\epsilon}+ {\rm{Ls}}_{-1}\bigg(\frac{-s_{13}}{-s_{123}},\frac{-s_{12}}{-s_{123}}\bigg)\nonumber\\&&+ {\rm{Ls}}_{-1}\bigg(\frac{-s_{23}}{-s_{123}},\frac{-s_{12}}{-s_{123}}\bigg)\bigg)+c_{\Gamma}\frac{s_{13}+s_{23}}{2\spb3.2\spb1.3}~.
\label{eq:s_virtsl}
\end{eqnarray}
Unlike the vector case of Appendix~\ref{app:VecC}, there is no Ward identity for the scalar operator.  This results in a further UV counter term in addition to those specified in Appendix~\ref{app:VecC}. For example, in the $\overline{\rm{MS}}$-scheme one must include the counterterm $-c_{\Gamma}\frac{3}{2\epsilon}\mathcal{A}^{(0)}_{S}$ in~(\ref{eq:s_virtlc})-(\ref{eq:s_virtsl}).
For the calculation of the real corrections we will need the following amplitudes which involve the emission of an additional parton with respect to the Born monojet process
\begin{eqnarray}
A^{(0)}_S(1^-_q,2^-_{\overline{q}},3^-_g,4_g^-)&=&\frac{s_{1234}}{\spb1.4\spb2.3\spb3.4}~, \\ 
A^{(0)}_S(1^-_q,2^-_{\overline{q}},3^+_g,4_g^+)&=&\frac{\spa1.2^2}{\spa1.4\spa2.3\spa3.4}~, \\ 
A^{(0)}_S(1^-_q,2^-_{\overline{q}},3^-_g,4_g^+)&=&\frac{\spa2.3\spb4.2\spab 1.(2+3).4}{s_{234}\spa3.4\spb2.3\spb3.4}+
\frac{\spa1.3^2\spab2.(1+3).4}{s_{134}\spa1.4\spa3.4\spb3.4}\nonumber\\&&-\frac{\spa1.3\spab1.(2+3).4}{\spa1.4\spa3.4\spb2.3\spb3.4}~, \\
A^{(0)}_S(1^-_q,2^-_{\overline{q}},3^+_g,4_g^-)&=&A^{(0)}_S(1^-_q,2^-_{\overline{q}},4^-_g,3_g^+)~.
\end{eqnarray}
In addition we will need the four-quark amplitude which is given by, 
\begin{eqnarray}
A^{(0)}_S(1^-_q,2^-_{\overline{q}},3^-_{Q},4_{\overline{Q}}^+)&=&\frac{\spa1.3\spab2.(1+3).4}{s_{134}\spa3.4\spb4.3}+
\frac{\spa2.3\spab1.(2+3).4}{s_{234}\spa3.4\spb4.3}. 
\end{eqnarray}
Finally for the case involving monophoton production the amplitudes we need are as follows, 
\begin{eqnarray}
A^{(0)}_S(1^-_q,2^-_{\overline{q}},3^-_g,4_{\gamma}^-)&=&\frac{s_{1234}\spb1.2}{\spb1.4\spb2.3\spb1.3\spb2.4}~, \\ 
A^{(0)}_S(1^-_q,2^-_{\overline{q}},3^+_g,4_{\gamma}^+)&=&\frac{\spa1.2^2}{\spa1.4\spa1.3\spa2.3\spa2.4}~, \\ 
A^{(0)}_S(1^-_q,2^-_{\overline{q}},3^+_g,4_{\gamma}^-)&=&\frac{\spa1.4\spab2.(1+4).3}{s_{134}\spa1.3\spb4.1}-\frac{\spa2.4\spab1.(2+4).3}{s_{234}\spa2.3\spb4.2}\nonumber\\&& +\frac{\spa1.2s_{124}}{\spa1.3\spa2.3\spb4.1\spb4.2}~, \\
A^{(0)}_S(1^-_q,2^-_{\overline{q}},3^-_g,4_{\gamma}^+)&=&A^{(0)}_S(1^-_q,2^-_{\overline{q}},3^+_g,4_{\gamma}^-) (3\leftrightarrow 4)~.
\end{eqnarray}

\subsection{SM Gluonic currents} 
The production of DM in association with a jet through the gluon operator 
\begin{eqnarray}
\mathcal{O}_g=\alpha_s\frac{(\chi\overline{\chi})(G^{\mu\nu}_aG_{a,\mu\nu})}{\Lambda^3}~,
\end{eqnarray}
is simple to implement due the large existing literature associated with Higgs boson production in the heavy-top effective theory. In fact, modulo changes in the coupling the process is identical 
to $H\rightarrow (b\overline{b})+$jet. Therefore we simply modify the existing MCFM routines accordingly~\cite{EHSB,Campbell:2010cz}. Clearly no monophoton signature is produced from this operator.

\section{Monojet phenomenology } \label{sec:monojet}

In this section we present some phenomenological studies using the results derived in the previous sections. We have implemented our NLO calculations into MCFM, taking full advantage of the codes existing architecture to obtain our predictions. More specifically, we use the dipole subtraction scheme of Catani and Seymour~\cite{Catani:1996vz} 
to render both the virtual and real corrections separately finite.  We use the following MCFM default 
electroweak (EW) parameters in our calculation, 
\begin{eqnarray}
M_Z = 91.1876 \,\,{\rm{GeV}} &\;, \quad & M_W = 80.398  \,\,{\rm{GeV}} \;,\nonumber\\
\Gamma_Z =2.4952  \,\,{\rm{GeV}}& \;, \quad &\Gamma_W = 2.1054 \,\,{\rm{GeV}} \;, \nonumber\\
G_F=0.116639  \times 10^{-4}  \,\,{\rm{GeV^{-2}}} &\;, \quad& m_t = 173.2 \,\,{\rm{GeV}} \;. \nonumber
\end{eqnarray}
The remaining EW parameters are defined using the above as input parameters. In order to avoid DM being charged under $SU(2)$ the scalar operator proceeds through a Yukawa coupling and is thus proportional to the quark masses. In these cases we use $m_c=1.5$, $m_b=4.7$ GeV and $m_t$ as defined above with all other quarks kept massless. For the light quarks $(m_c$ and $m_b$) the masses are only retained in the form of a Yukawa coupling, and are not retained in the kinematics of the matrix element. For a pure monojet analysis this is of little consequence.  In the case of the top quark, which can contribute to monojet processes at the loop level, the mass is fully retained.  Our default PDF choice is CTEQ6L1 for LO and CT10 for NLO calculations~\cite{Lai:2010vv}. 

In this section we will use the effective theory prescription defined 
earlier. These EFTs are adequate descriptions of a more UV complete model, provided the mediator scale $\Lambda$ is large enough. Our implementation of the various operators into MCFM allows for flavour dependent couplings to be used, so that up- and down-type operators can be studied individually if so required. In these examples we will consider the operators to be flavour diagonal, coupling to each flavour with the same coupling, taken to be $1$. The only exception to this rule are the operators $\mathcal{O}_S$ and $\mathcal{O}_{PS}$, which as discussed above couple with quark-mass dependent couplings. 
 
 \subsection{Monojet inclusive cross sections}
 
We now turn to phenomenology, we begin by considering the production of a pair of DM particles in association with a jet.  In this section we focus on the LHC operating at 7 TeV and we will use cuts inspired by the recent ATLAS and CMS publications~\cite{ATLAS:2012ky,Chatrchyan:2012me} requiring events to pass the following phase space cuts, 
\begin{eqnarray} 
\slashed{E}_T > 350 \,\,{\rm{GeV}}\,, \; \quad p^j_{T} > 100 \,\,{\rm{GeV}}\,,  \; \quad|\eta_j| < 2\,, \quad \Delta\phi_{j_1,j_2} < 2.5~.
\label{eq:cuts}
\end{eqnarray}
Jets are defined using the anti-$k_T$ algorithm with $R=0.4$.
Experimental searches also implement a veto on more than two jets, however this kinematic configuration is not covered by the NLO calculation and 
as such this cut is meaningless in our studies.  As discussed earlier, at LO the mismatch in the $\met$ and jet-$p_T$ requirements means that only events with the jet $p_T$  above the $\met$ cut will be included.  At NLO there is a new region of phase space with a jet and an additional parton (which may or may not pass the jet algorithm) each with a $p_T$ below the $\met$ cut, that can conspire to produce enough $\met$ so that the event passes all cuts.  We note that the cut on the azimuthal separation between the two jets $\Delta\phi_{j_1,j_2}$ only enters our calculation at NLO and as such is a fairly weak cut. 

\begin{figure}[t]
\includegraphics[width=8cm]{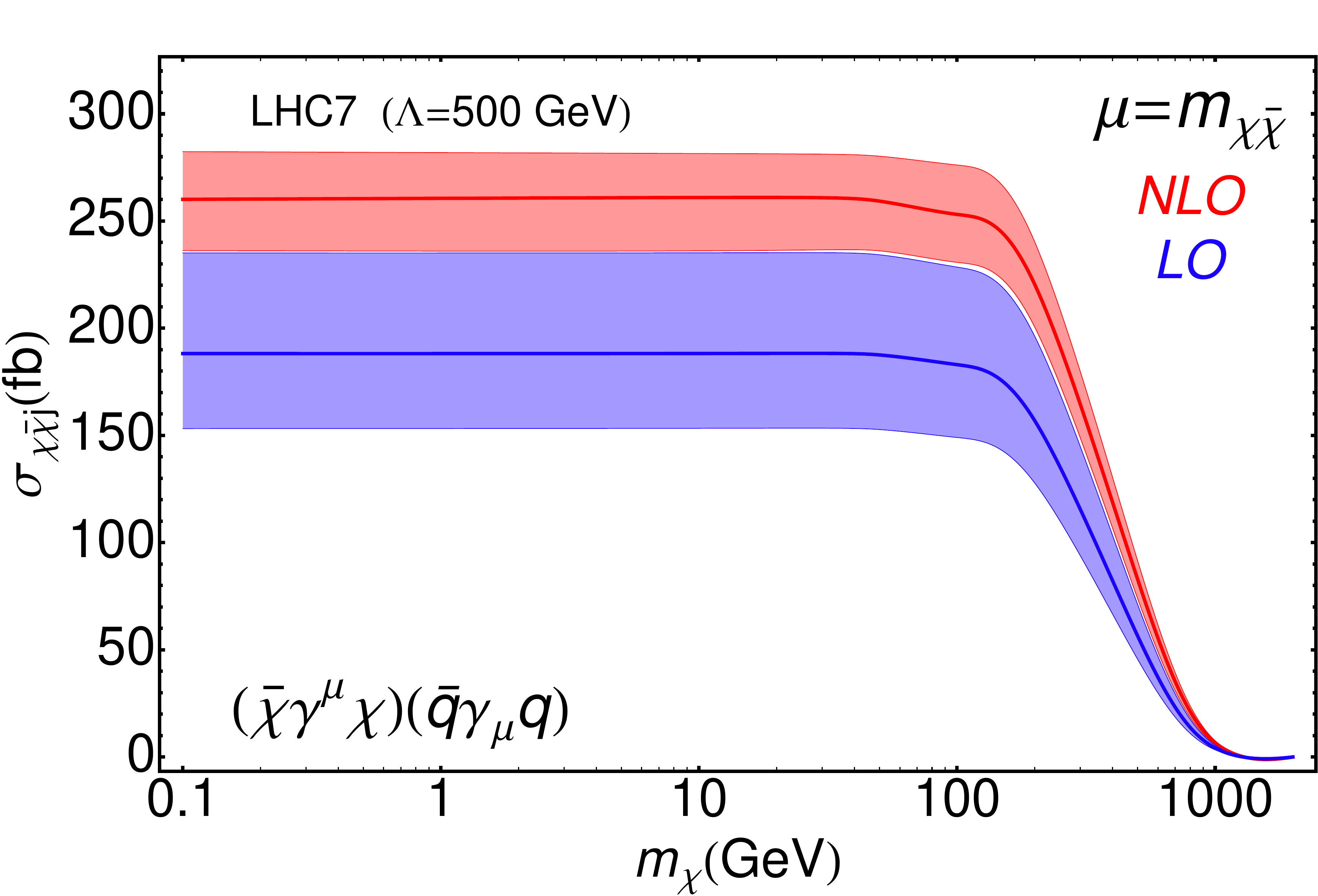}
\includegraphics[width=8cm]{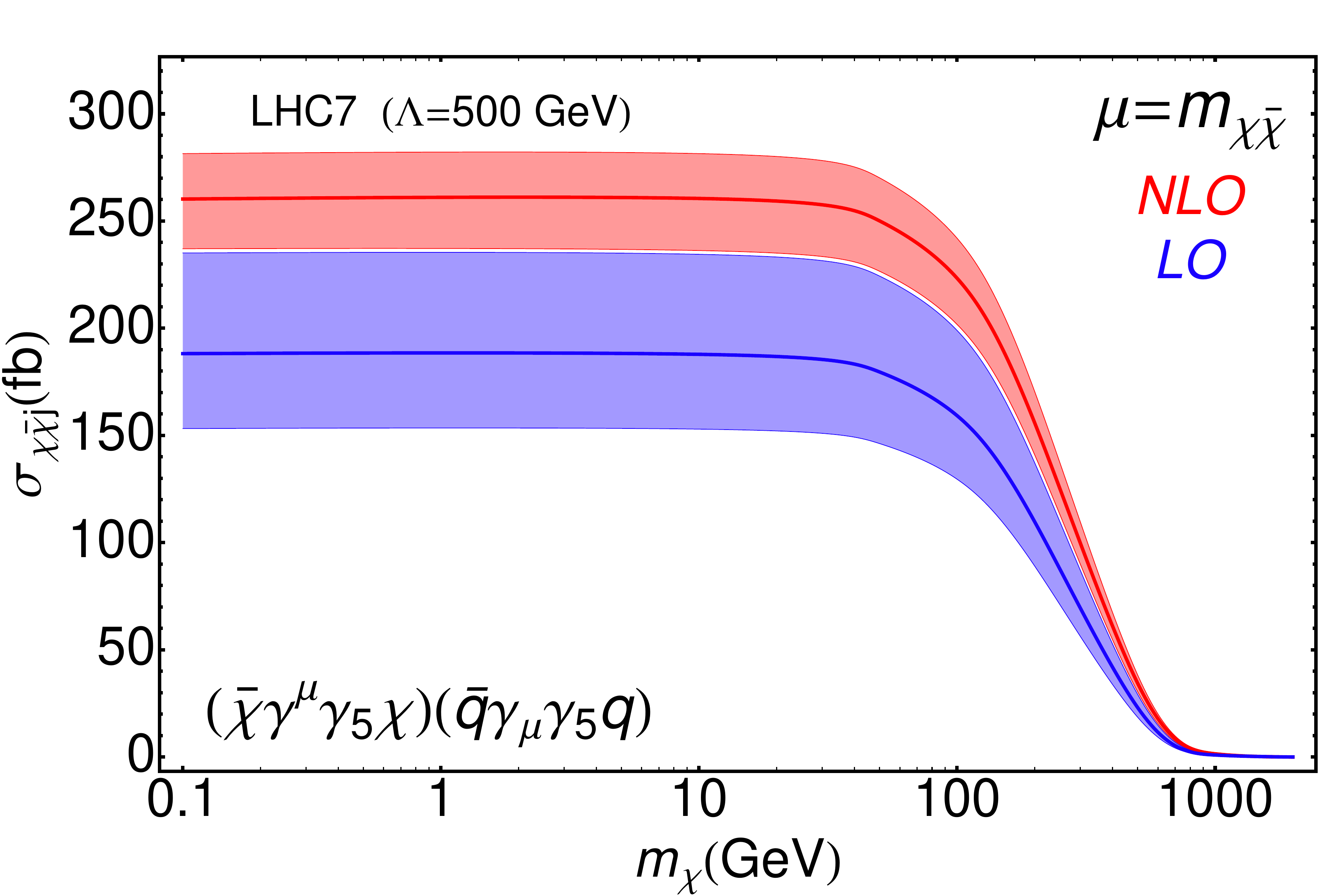}
\begin{center} 
\end{center}
\includegraphics[width=8cm]{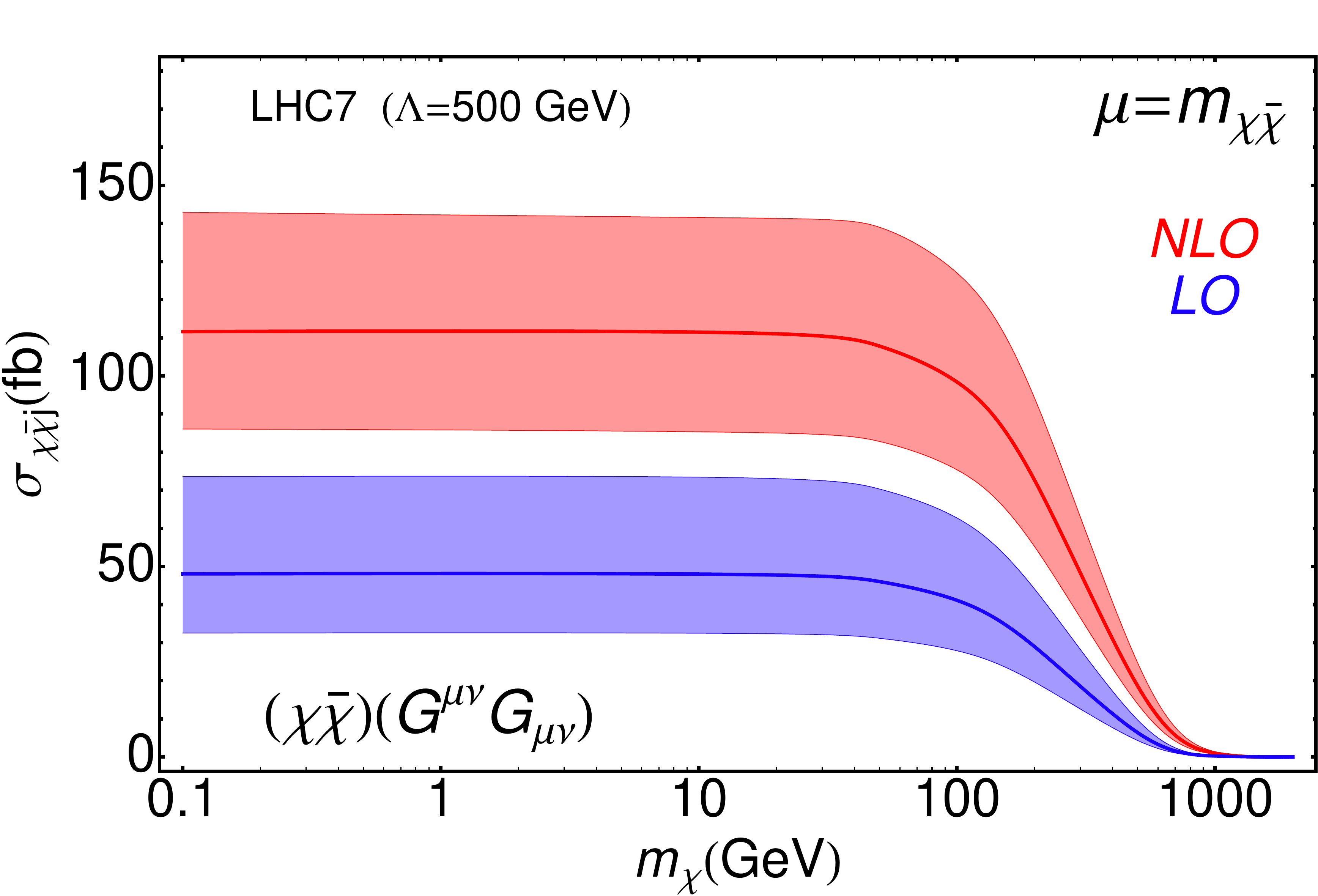}
\includegraphics[width=8cm]{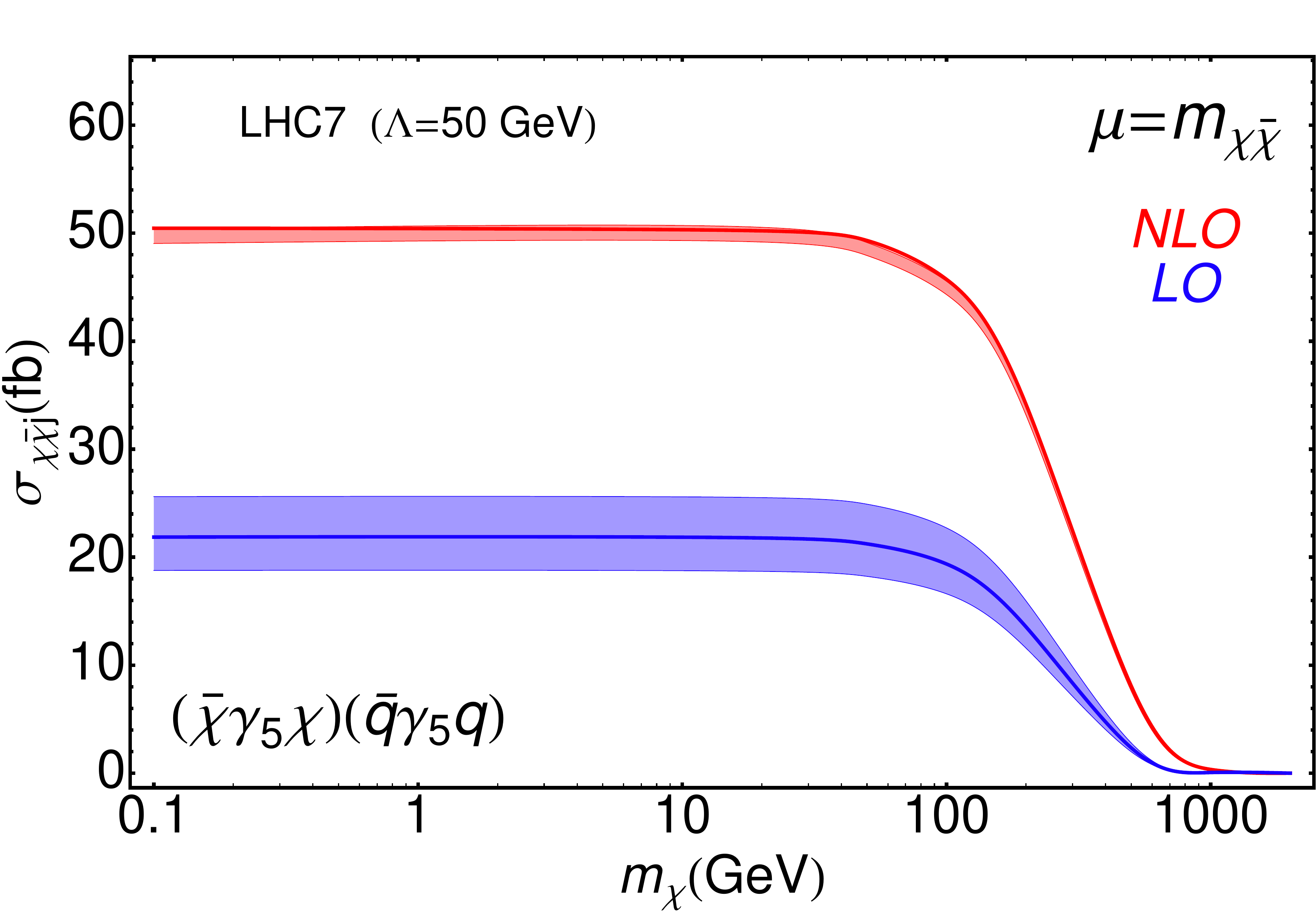}
\begin{center} 
\includegraphics[width=8cm]{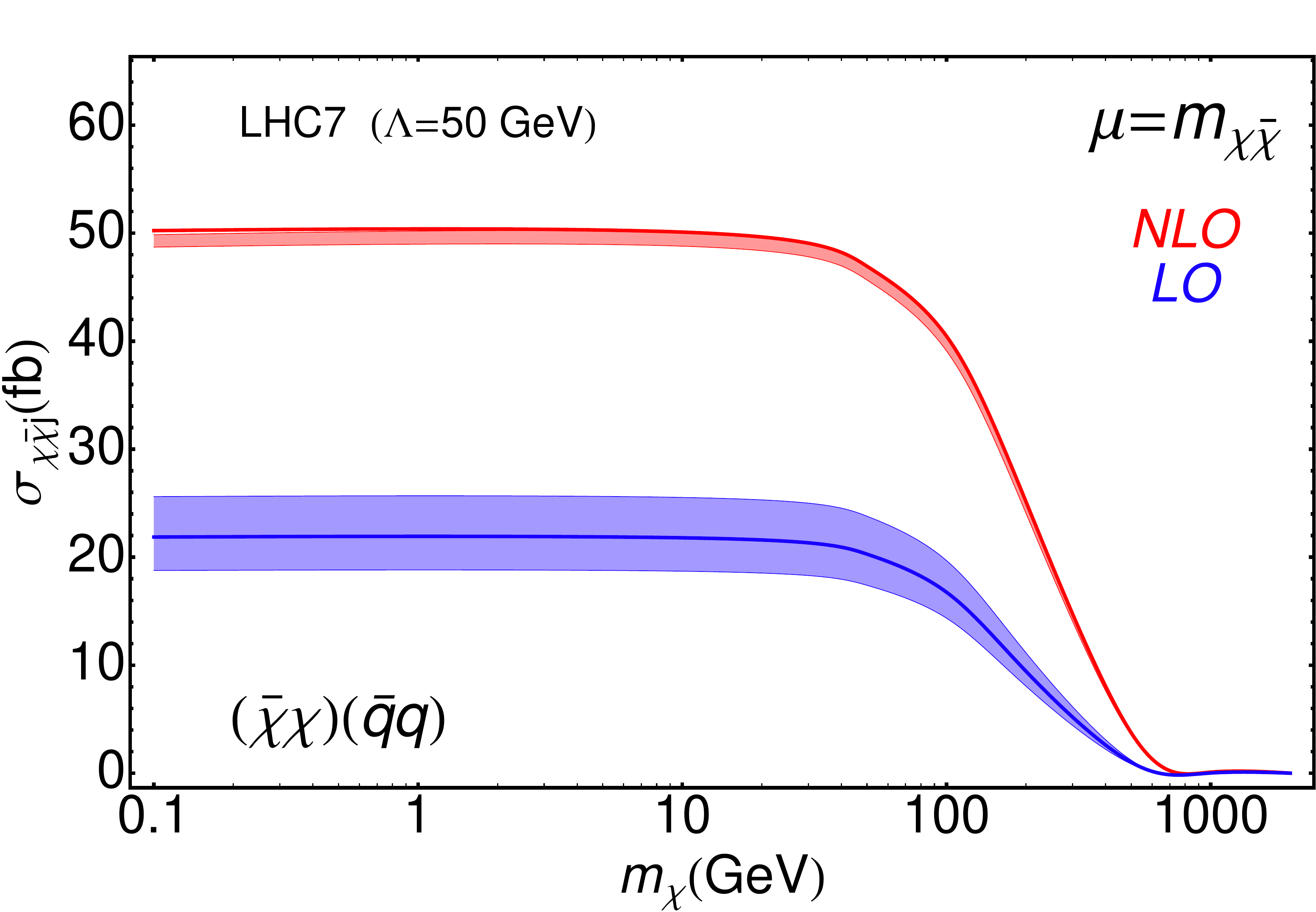} 
\includegraphics[width=8cm]{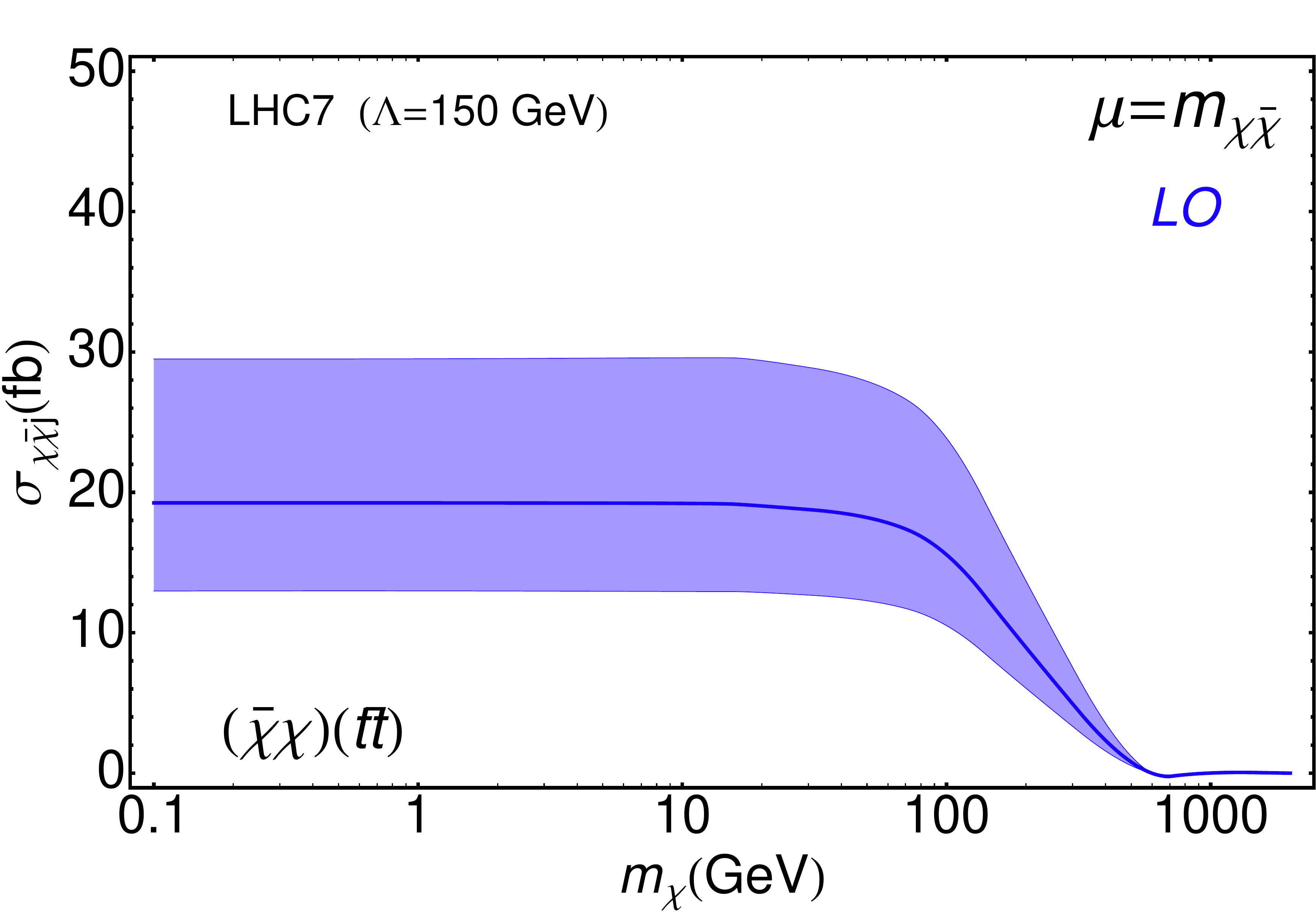} 
\end{center}
\caption{LO and NLO cross sections for DM production in association with a jet at the 7 TeV LHC. The solid line indicates the cross section obtained with the default scale $\mu=m_{\chi\overline{\chi}}$, the shaded band represents the deviation from this scale when the scales are varied by a factor of two in each direction. The phase space cuts described in the text (\ref{eq:cuts}) have been applied.}
\label{fig:xs_monoj_r2}
\end{figure}

\begin{figure}[t]
\includegraphics[width=8cm]{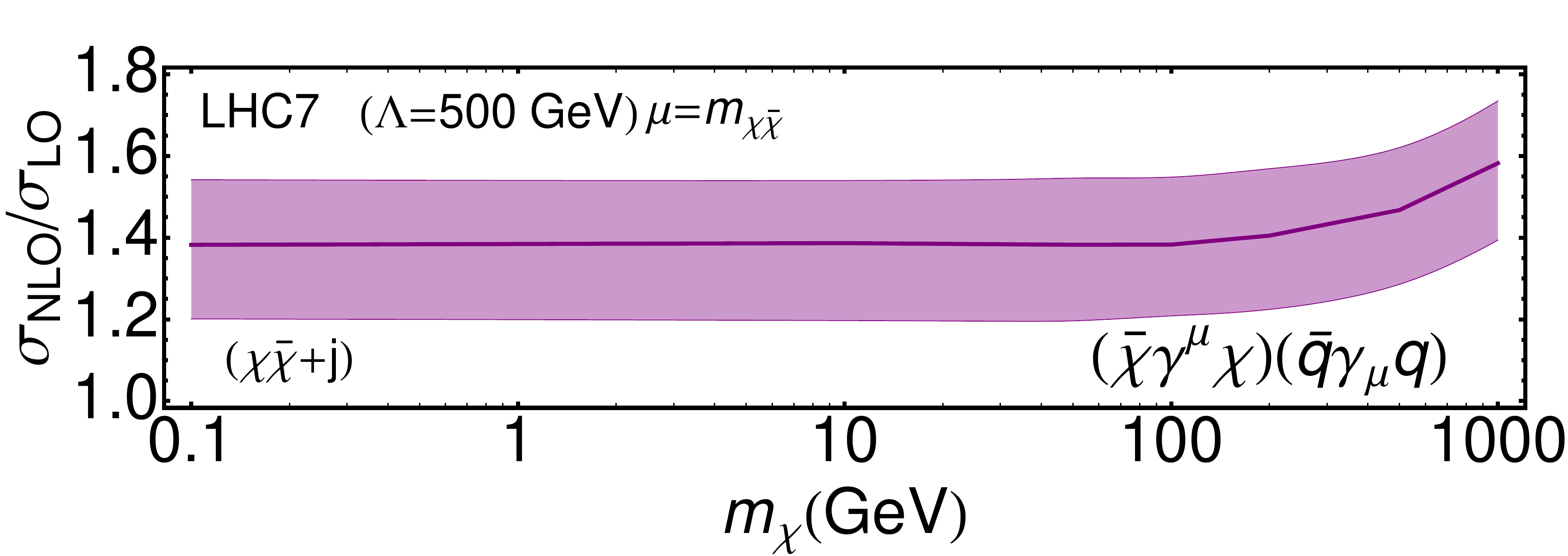}
\includegraphics[width=8cm]{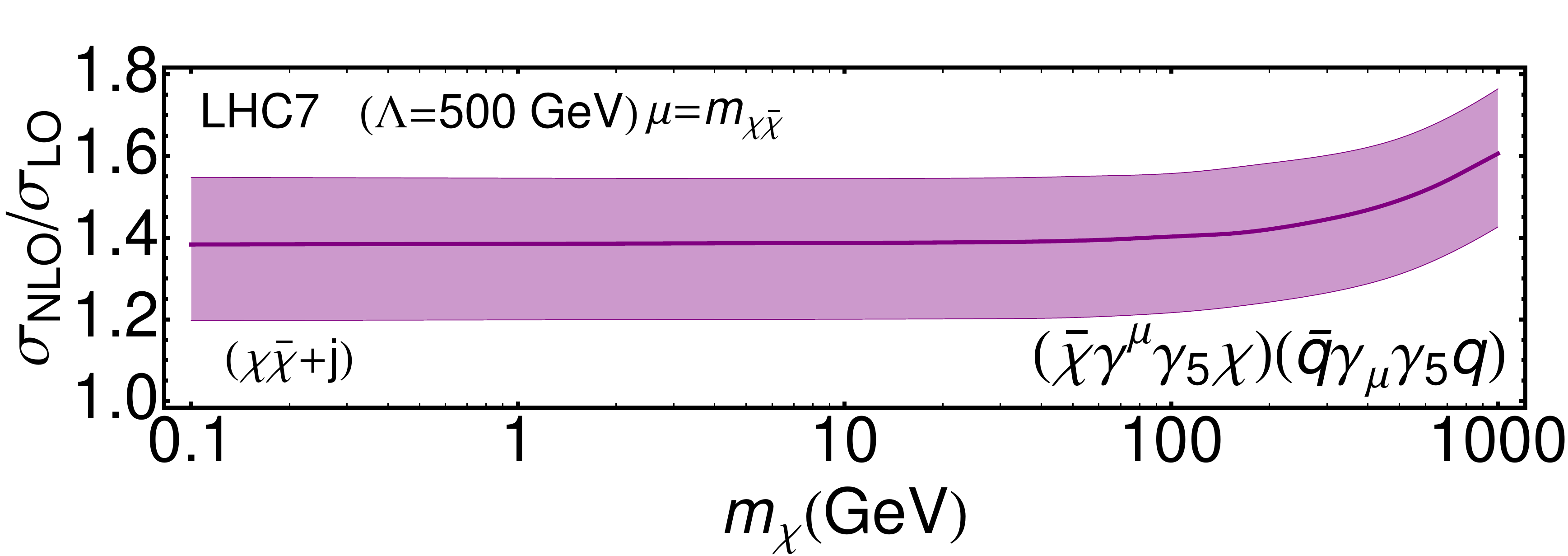}
\begin{center} 
\end{center}
\includegraphics[width=8cm]{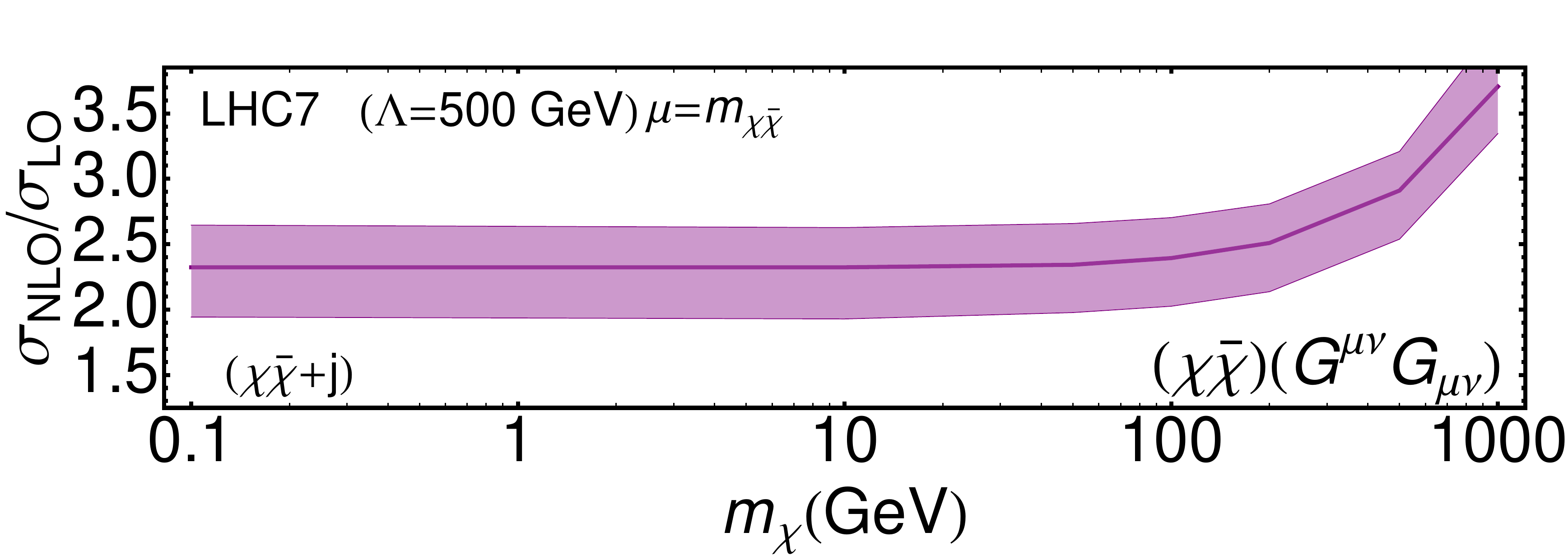}
\includegraphics[width=8cm]{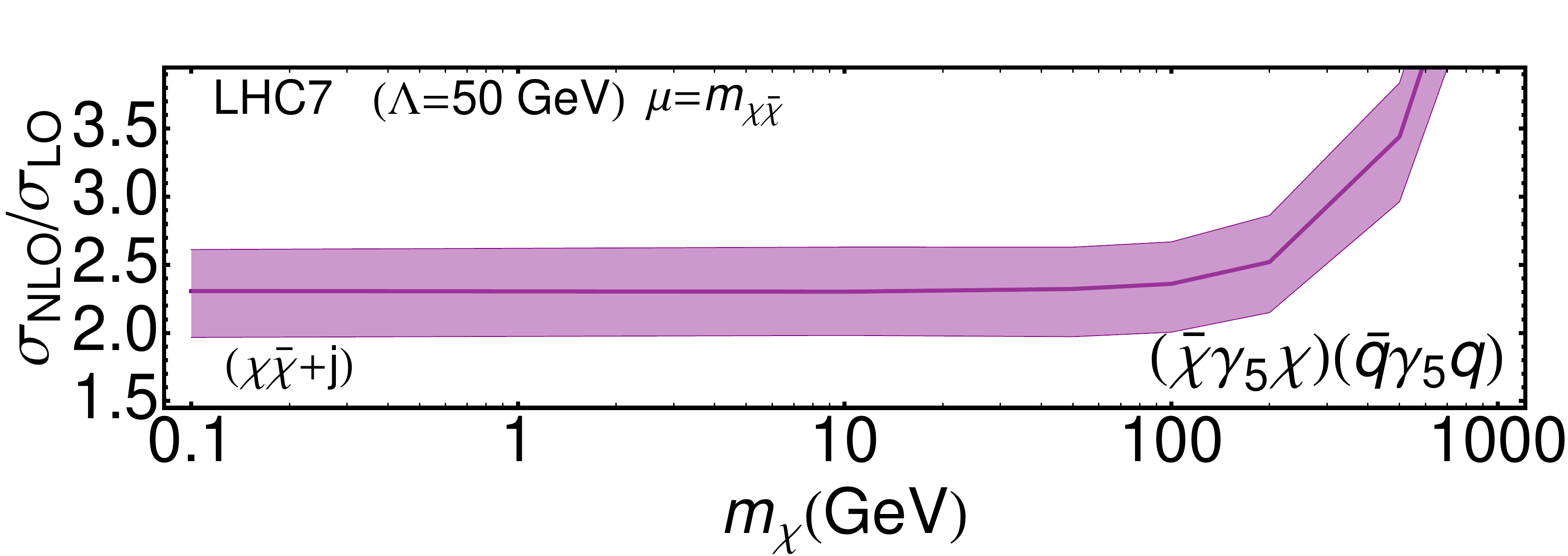}
\begin{center} 
\includegraphics[width=8cm]{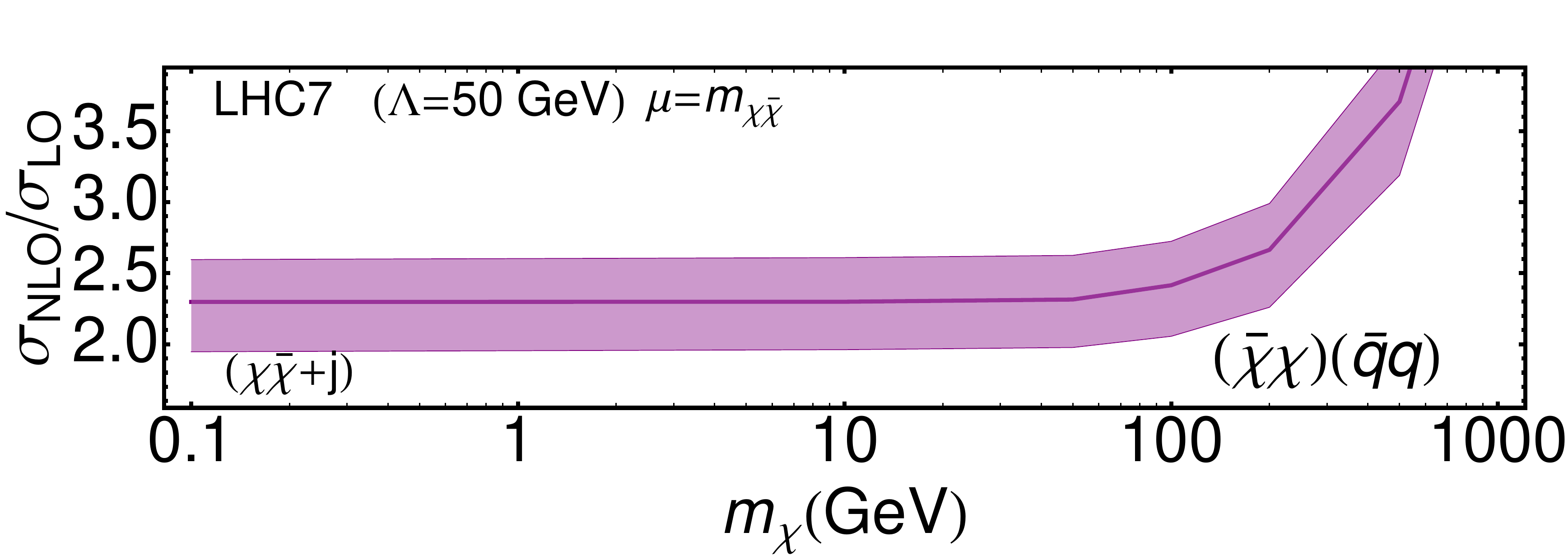} \\
\end{center}
\caption{$K$-factors for DM production in association with a jet at the 7 TeV LHC. The solid line indicates the $K$-factor obtained with the default scale $\mu=m_{\chi\overline{\chi}}$, the shaded band represents the $K$-factors obtained using scales varied by a factor of two in each direction. The phase space cuts described in the text (\ref{eq:cuts}) have been applied.}
\label{fig:K_monoj_r2}
\end{figure}

Using these cuts we begin by studying total inclusive cross sections and their inherent scale dependence at LO and NLO. Our results are summarized in Fig.~\ref{fig:xs_monoj_r2}.  We have chosen the values of $\Lambda$ for each operator to get close to the rate the experiments are presently placing a bound on, $\mathcal{O}(10-100)\,\fb$.  We have made no attempt to actually place limits directly on $\Lambda$. In addition we provide the ratio of NLO to LO cross sections (the $K$-factor) for each operator in Fig.~\ref{fig:K_monoj_r2}. The dependence on 
the renormalization and factorization scales, taken to be equal ($\mu_R=\mu_F=\mu$), is illustrated by the shaded band linking the predictions obtained at $\mu=2m_{\chi\overline{\chi}}$ and $\mu=1/2m_{\chi\overline{\chi}}$, whilst the central scale choice $\mu=m_{\chi\overline{\chi}}$ is illustrated by the curve inside the shaded band. From the various curves it is clear that the NLO corrections to the total cross section are sizeable, and vary operator to operator. As is natural by going to NLO the overall scale dependence is smaller than the LO prediction. Before we comment on the results for each DM operator individually we observe that all the plots have a similar shape. Namely a flat plateau in the light $m_{\chi}$ region in which the cross section is insensitive to $m_{\chi}$, and then a rapid fall off at a DM mass of around $100$ GeV. These features have been observed in previous phenomenological studies~\cite{Beltran:2010ww,Goodman:2010yf,Goodman:2010ku,Rajaraman:2011wf,Fox:2011pm,Shoemaker:2011vi,Fortin:2011hv} and indeed are clear in the experimental constraints on $\Lambda$~\cite{ATLAS:2012ky,Chatrchyan:2012me}. For light DM the analysis cuts are sufficiently hard that a $\sqrt{s}=7$ TeV collider sees no difference in terms of phase space restrictions in producing DM with mass 10 GeV or 0.1 GeV.  This explains the plateau in production cross section as a function of DM mass. Then as the DM mass becomes a non-negligible scale in the process the phase space restrictions begin to rapidly drive the cross section downwards, resulting in the distinctive fall off in each of the plots.   The point at which this fall off begins depends on the exact mass dependence of the cross section and is therefore operator specific.  As the operating energy increases, the fall off moves to larger $m_{\chi}$, for every operator.  As such, the limits on $\Lambda$ obtained using the 8 TeV data set should show a considerable improvement (beyond the simple $\sqrt{s}$ rescaling) for DM masses around 0.1-1 TeV. 

We now examine each operator in detail, firstly we observe that the vector and axial-vector DM operators show similar behaviour in terms of $K$-factors and scale dependence. This is unsurprising since in the massless limit the only terms which are sensitive to the axial nature of the coupling are the four-quark amplitudes, which are a small part of the total NLO cross section.  As a result it is clear that as $m_{\chi} \rightarrow 0$ the results must be similar.   As the the DM mass grows, so does the difference between the operators, with the axial operator being smaller than the vector over the entire mass range.  The scale variation for these operators is also similar, with typical LO values around $\pm 20\%$ and NLO values around $\pm 10\%$. The $K$-factor for both these operators is around 1.4.  Since, for this operator, the cross section scales as $\Lambda^{-4}$ one naively expects an improvement on the limits on $\Lambda$ of around a $~\sim 10 \%$ if the NLO rate were used compared to the LO one. 

The $\mathcal{O}_g$ operator shows some striking differences with respect to the  $\mathcal{O}_V$ and  $\mathcal{O}_A$ operators. Firstly the $K$-factors are much larger, with values between 2 and 2.5, indicating much larger NLO corrections for this operator. Secondly, the scale dependence is large at both orders. The LO scale dependence is around 50\% reducing to around 25\% at NLO. These effects are reminiscent of the Higgs effective operator, in which large scale dependence and $K$-factors are common. However, there is a crucial difference between the two calculations which should not be overlooked, for the case of a light Higgs the final state phase space is extremely restricted by the narrow $s$-channel resonance. In our effective theory, no such restriction applies, allowing for the possibility of enhancing the NLO effects by sampling over a larger phase space. Since the operator scales as $\Lambda^{-3}$ using a NLO cross section to set a limit on $\Lambda$ should result in improvements of the order of $\sim 16\%$.

The scalar and pseudo-scalar operators (for $m_b$ and $m_c$ couplings) ($\mathcal{O}^{(c,b)}_S$ and $\mathcal{O}^{(c,b)}_{PS}$) behave in a rather different manner to the other operators. They are naturally suppressed by the factor $m_q/\Lambda$ compared to the vector, and axial cases, which results in much smaller cross sections. For this reason we have chosen to use a smaller value of $\Lambda$ in our analysis of these operators (50 GeV compared to 500 GeV for the other cases). The LO diagrams require the presence of a heavy sea quark (charm or bottom) in order to be non-zero, therefore it is natural to expect a large $K$-factor at NLO arising from contributions in which the process is initiated by two gluons. This is indeed the case, we observe a $K$-factor of around  2-2.5, leading to an improvement in the $\Lambda$ bound of $\sim 25\%$. The scale dependence is markedly reduced at NLO compared to LO, (around $5\%$ at NLO compared to around $20\%$ at LO). However, given the large $K$-factor going from LO to NLO it is unlikely that scale variation alone gives a reliable estimation of theoretical uncertainty (even at NLO) for this operator. 

Finally, we consider the production of a monojet signature using the  $\mathcal{O}^{(t)}_S$ operator.  The enhancement which comes from including the loop induced processes in which the DM couples directly to the top quarks, is very large~\cite{Haisch:2012kf}. We show the LO cross sections obtained using MCFM where for this operator we have set $\Lambda=150$. The cross sections obtained using the top-induced operator is around two orders of magnitude larger than those for the light quarks. As a result if this operator was used in place of the light quark operator the corresponding limit on would increase $\Lambda$ by between 200 and 300\%~\cite{Haisch:2012kf}. Since this processes is LO (even though it is a loop induced process) the scale dependence is very large, around (40-50\%). This means that a sensible strategy experimentally may be to set limits on the operators induced by top and bottom couplings individually. Although the bottom induced $\Lambda$ will be much smaller than that for the top induced coupling it will suffer from much smaller systematic uncertainties on the theoretical side, since the NLO prediction can be used. However, we stress that the resulting $\Lambda$ limit obtained by either operator is unlikely to be within the realm of validity for the effective theory. For the top-scalar operator the limit is salvageable by switching to the full theory as we will investigate in the next section.  The rate for the light-quark operators are suppressed by quark mass and so the EFT approach is brought into question.  In  a simple UV completion of a scalar mixing with the Higgs, discussed below, the rate is also expected to be very small.  However if a, naively, tuned model is considered in which the DM couplings are diagonal in the quark mass basis the couplings could instead all be $\mathcal{O}(1)$ and these operators would have non-trivial bounds.  Given that the production of DM through the scalar operators probes a different set of PDFs from the vector operators it is still worth investigating the scalar operator in general.

\begin{figure}
\includegraphics[width=8cm]{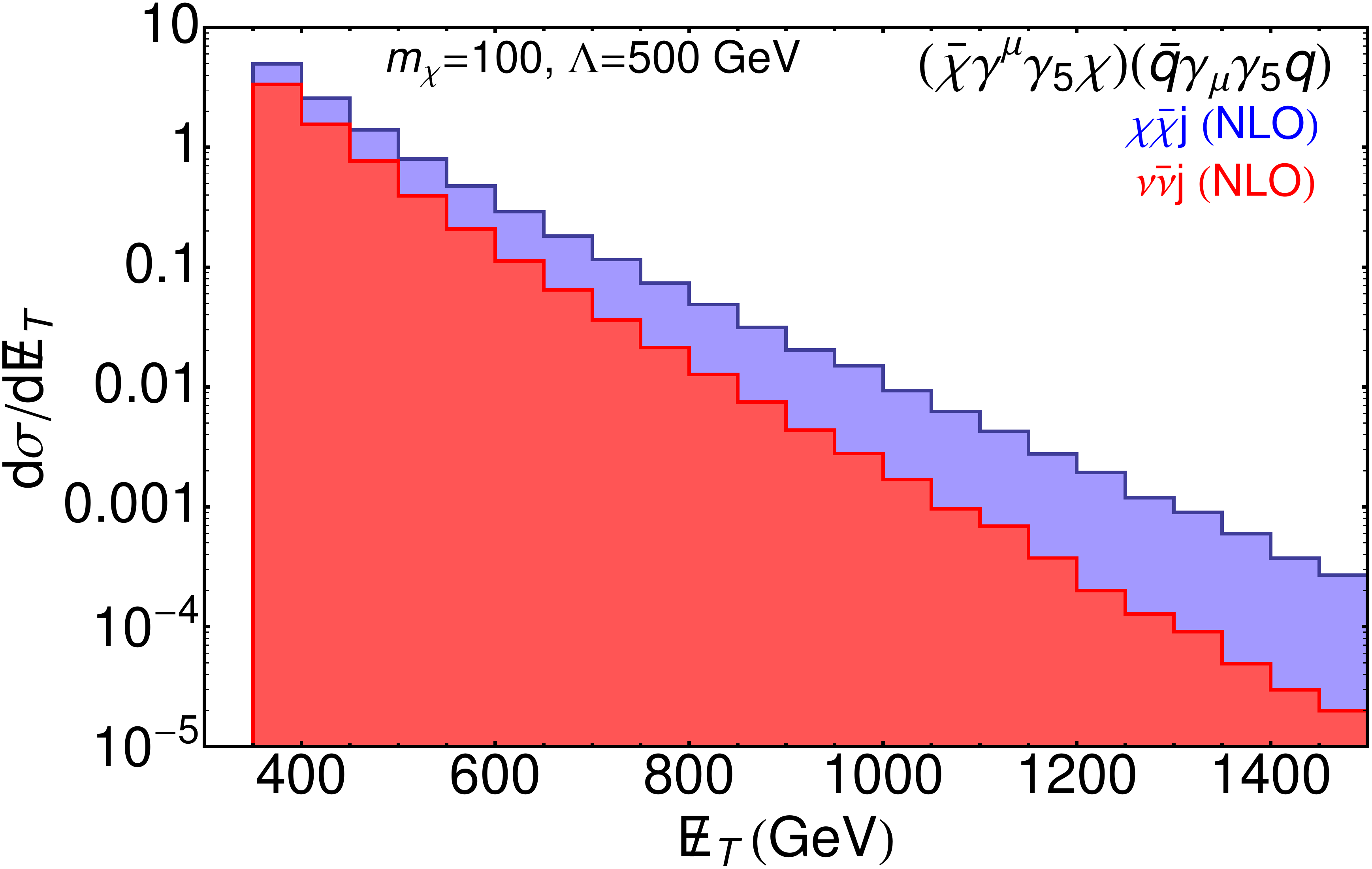}
\includegraphics[width=8cm]{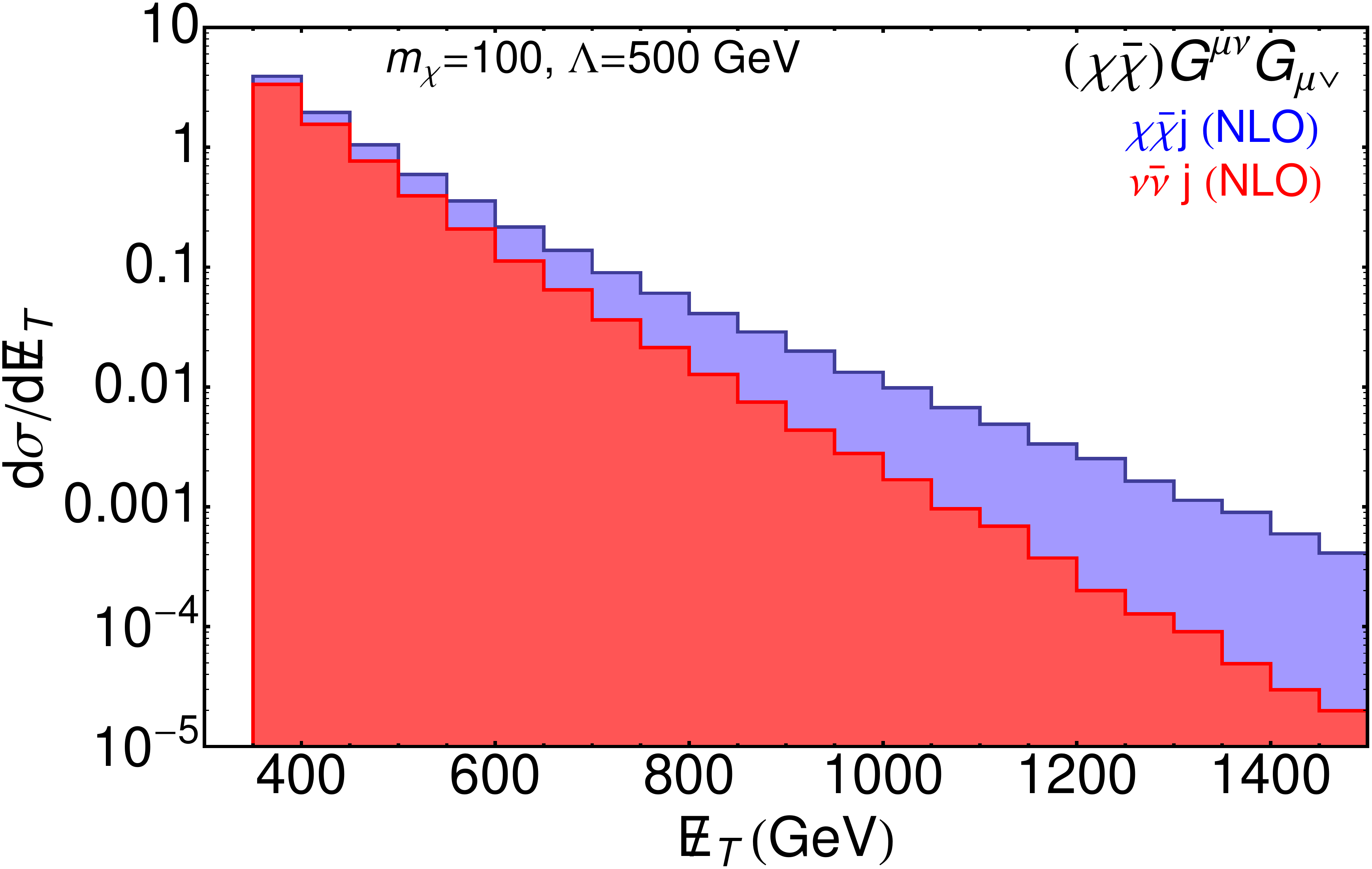}
\caption{NLO predictions for the missing transverse momentum spectrum for signal and background obtained using MCFM. 
We present results for two operators ($\mathcal{O}_A$ and $\mathcal{O}_{g}$.) The DM mass is 100 GeV, and the scales have been chosen to be equal to the DM invariant mass $\mu=m_{\chi\overline{\chi}}$.}
\label{fig:MET_mj}
\end{figure}

 \subsection{Monojet differential distributions}

\begin{figure}
\includegraphics[width=8cm]{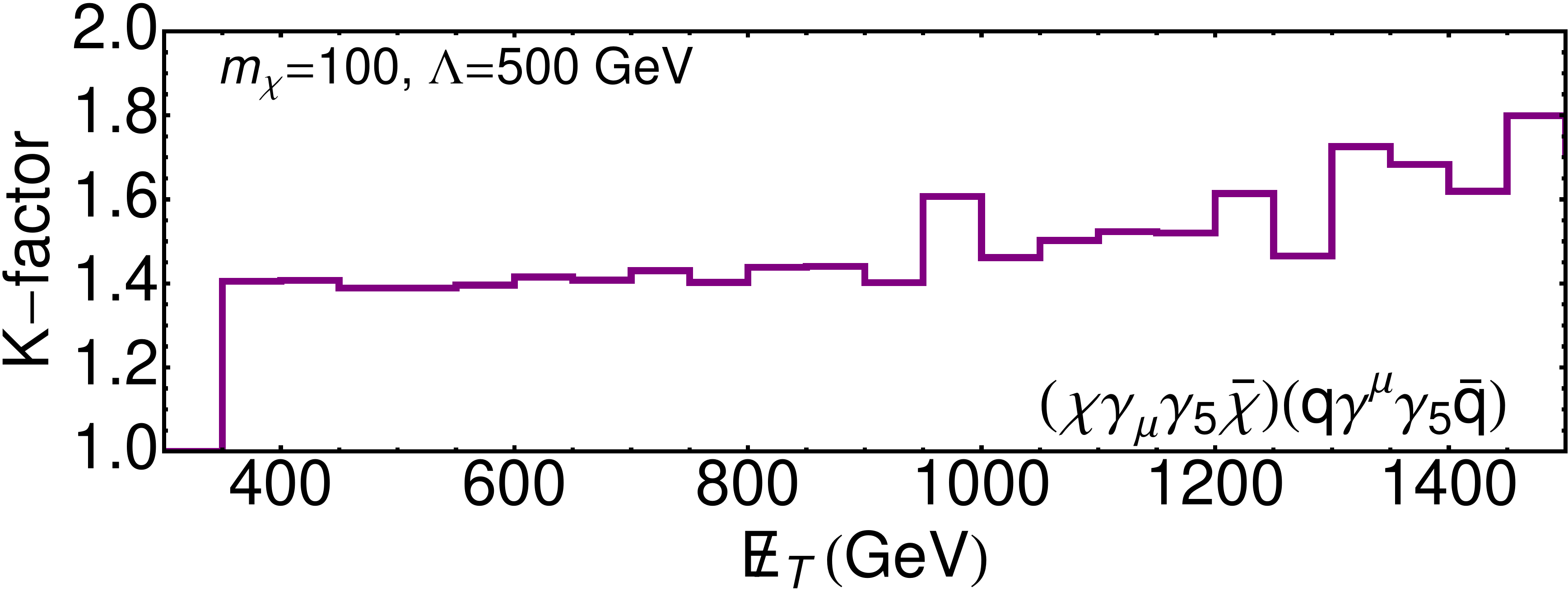}
\includegraphics[width=8cm]{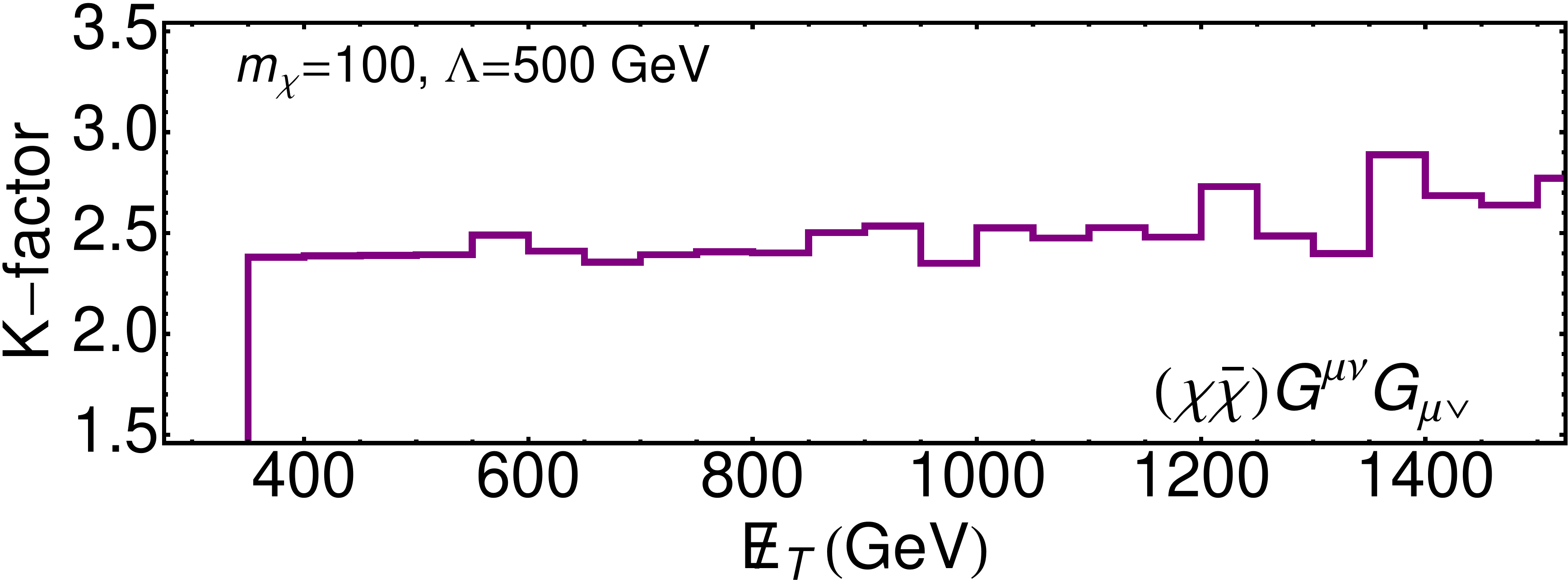}
\caption{The differential $K$-factor, $(d\sigma_{NLO}/d\met)\big/(d\sigma_{LO}/d\met)$, for monojet production via either an axial or gluon-induced operator for $m_{\chi}=100$ GeV.}
\label{fig:MET_K}
\end{figure}

So far we have focussed our attention on predictions for total inclusive cross sections at NLO. It is also interesting 
to consider the NLO corrections to important distributions used in the experimental searches. As an example of such 
a distribution we consider the missing transverse momentum spectrum $\slashed{E}_T$. For simplicity we focus on 
two operators ($\mathcal{O}_A$ and $\mathcal{O}_g$) for a fixed DM mass of 100 GeV. We also use MCFM to obtain the spectrum for the dominant background contribution $Z\rightarrow(\nu\overline{\nu}) j$, our results are shown in 
Fig.~\ref{fig:MET_mj}. As was the case at LO the $\met$ spectrum for the DM signal is noticeably harder than for the that 
of the dominant background $Zj$. This can be understood by the scaling properties of the two spectrums, the signal to background ratio scales as $\Lambda^4/s^2_{\bar{\nu}\nu}$. Therefore, in the limit of large MET (large $s_{\bar{\nu}\nu}$), the EFT produces a harder spectrum than the background. This effect is dominated by the differences between the full and effective theory and as such is unchanged at NLO. In order to quantify the differences between LO and NLO for the signal distribution we present the differential $K$-factor in $\met$ in Fig.~\ref{fig:MET_K}. The choice of a dynamic scale at NLO results in a fairly stable differential $K$-factor with only a small growth with increasing $\slashed{E}_T$. 

\subsection{Light Mediators: Full theory considerations.}

The results presented above were obtained in the effective theory in which the particle responsible for mediating 
the interaction between the SM and the DM was taken to be heavy and was integrated out. It is interesting to consider the accuracy, the phenomenological validity, and the implications if this assumption breaks down. In particular, we note that, for the scalar and pseudo-scalar operators, in order to achieve DM production cross section of the size necessary to be observed at the LHC we must consider $\Lambda$ of $\mathcal{O}(50-200)$ GeV, see Fig.\ref{fig:xs_monoj_r2}.  Given the typical scales involved at the LHC, $\Lambda$ of this order cannot correspond to a mediator that can reasonably taken to be heavy, without entering the realm of strong coupling. As a result, we investigate the monojet phenomenology by reinstating the $s$-channel mediator. For simplicity we focus on two simple UV completions, the first being in which the mediator is a massive gauge boson which couples axially to the SM and DM fermions (\ie\ a straightforward UV completion of $\mathcal{O}_A)$.  The scale in the effective operator (\ref{eq:OA}) is related to the scale of the mediator by $\Lambda=M_\phi/\sqrt{g_q g_\chi}$ where the coupling to quarks (DM) is $g_q$ $(g_\chi)$. 

Secondly, we consider a UV completion of the scalar operator for top quarks only, $\overline{\chi}\chi \overline{t}t$.  A simple example of such a UV completion is to consider a new heavy singlet scalar $\Phi$ that mixes with the Higgs through a $\mu H^2 \Phi$ operator.  This leads, after electroweak symmetry breaking, to a coupling of the top to the mostly $\Phi$ mass eigenstate that is proportional to the top Yukawa.  If this heavy singlet also couples to DM as $g_\chi \Phi\bar{\chi}\chi$ then the amplitude for DM coupling rescales as
\be
\frac{m_t}{\Lambda^3}\rightarrow \frac{g_\chi y_t}{m^2_{\bar{\chi}\chi}-M_\phi^2+i M_\Phi\Gamma_\Phi}~.
\ee
Thus, the relationship between the full theory and the effective theory is more complicated and $\Lambda=(M_\phi^2 v/g_t g_\chi)^{1/3}$.

\begin{figure}
\includegraphics[width=8cm]{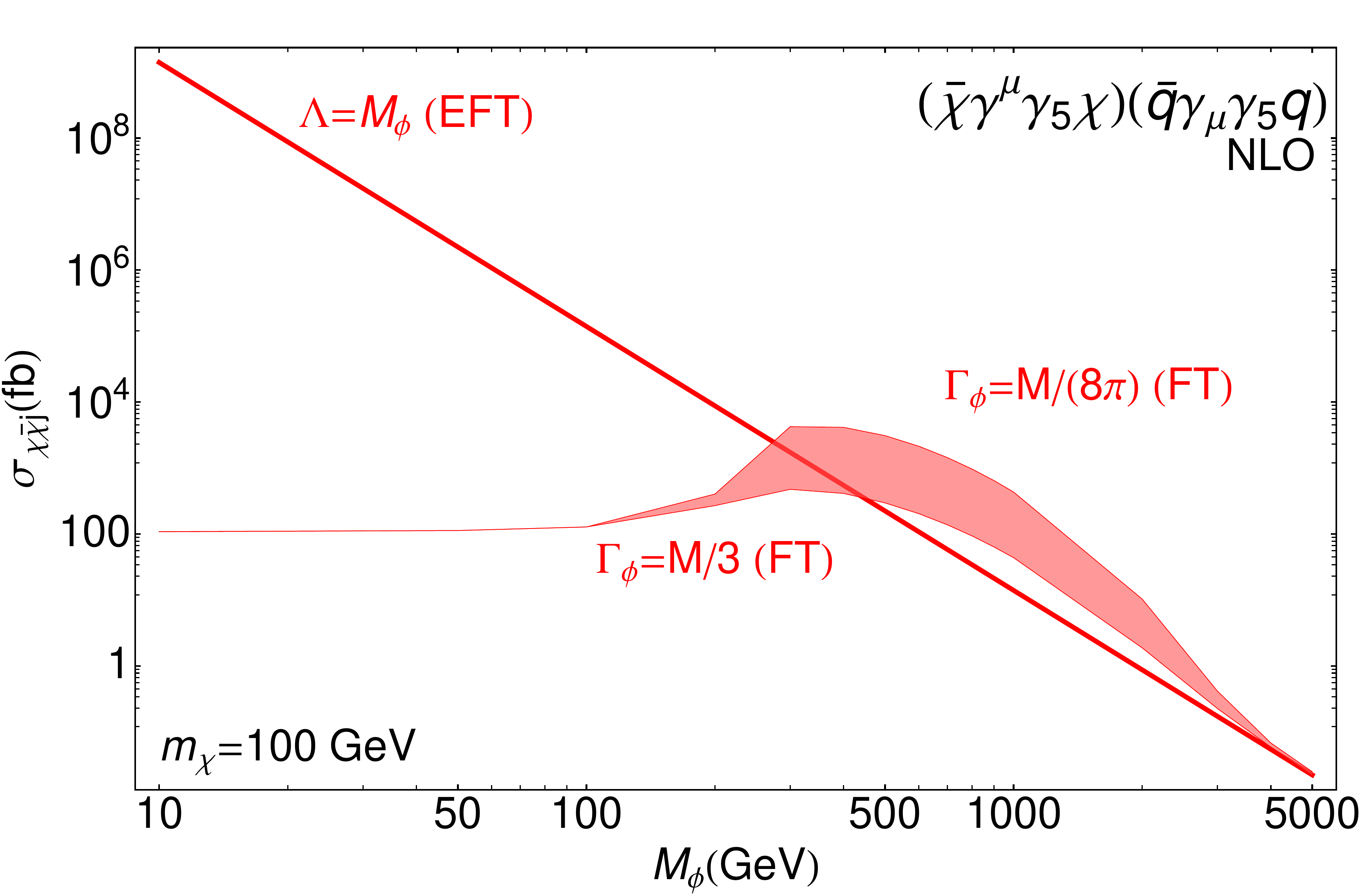}
\includegraphics[width=8cm]{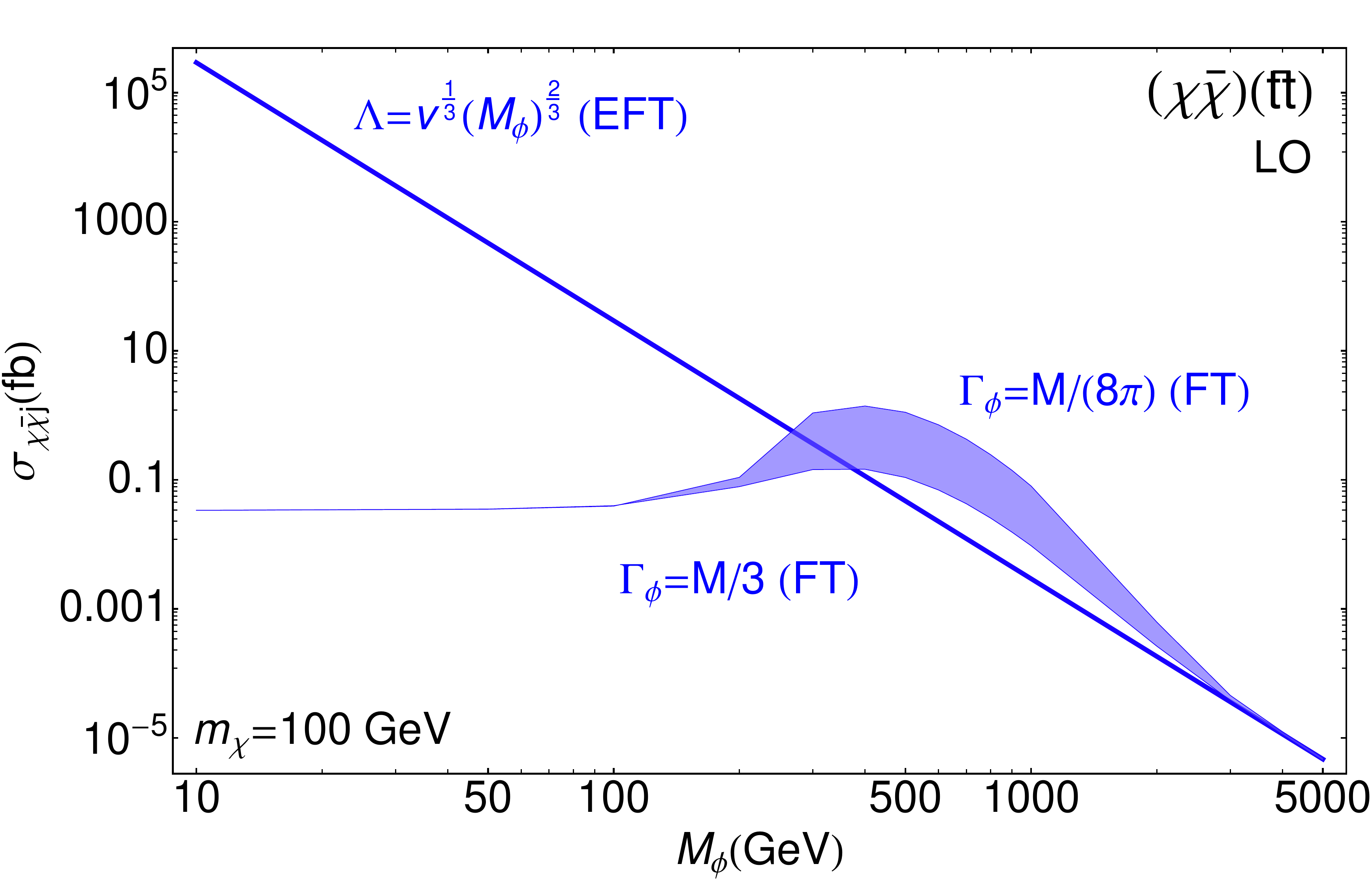}
\caption{Comparison between the full and effective field theories for axial and top-scalar operators, for DM mass of 100 GeV.  The straight line indicates the 
results obtained in the EFT, shaded region indicates full theory results as a function of the width between the values $M_{\phi}/3$ (lower) and $M_{\phi}/(8\pi)$ (upper). We have used the same phase space cuts as in the previous sections. The scale choice is $\mu=m_{\chi\overline{\chi}}$.}
\label{fig:FT_res}
\end{figure}
Results for these two theories are show in Fig.~\ref{fig:FT_res}, we have used the same final state phase space cuts as in the previous section and to aid in comparisons to the EFT we have fixed $g_{\chi}=g_{t}=g_{q}=1$, relating the scales of the operators in the EFT approach to the scale of the mediator as above. 
We then proceed to calculate cross sections as a function of $M_\phi$ in the full and effective theories. For the full theory one must also specify the width of the mediating particle $\Gamma_{\phi}$. We choose two example widths, $\Gamma_{\Phi}=M_{\phi}/3$ and $\Gamma_{\Phi}=M_{\phi}/(8\pi)$. The results are naturally dependent on the choice of DM mass, for simplicity we have chosen to focus on a single value $m_{\chi}=100$ GeV.

The two plots for the two different operators shown in Fig.~\ref{fig:FT_res} have some generic features which we explain first before mentioning some operator specific phenomenology. Firstly, it is clear that full theory asymptotes to the effective theory at large $M_\phi$ as required. On our log-log plot the EFT results then possess simple scaling as a function of $\Lambda$, and therefore $M_\phi$ (in the axial case) and $M_\phi^{2/3}$ (in the scalar case). 
All of the features of the FT are dominated by the inclusion of the propagator in the cross section, and in particular whether or not the propagator is able to provide resonant enhancement to the cross section. If the propagator is able to go on-shell then the cross section is enhanced beyond the EFT approximation, whereas if the propagator is forced into the off-shell region the cross section is suppressed (dramatically for light mediators) relative to the EFT.  

It is simple to determine whether the resonant enhancement will be included as a function of $m_{\chi}$ and $\slashed{E}_T^{min}$,
since, in order to achieve the on-shell condition one must have,
\begin{eqnarray} 
s_{\chi\overline{\chi}} \sim M_{\phi}^2 \implies (\slashed{E}_T^{min})^2+4m_{\chi}^2 <  M_{\phi}^2\,. 
\end{eqnarray}
Therefore if the $\met$ cut is too hard, making $\slashed{E}_T^{min}$ too large, or $m_{\chi}$ is too heavy then the mediator cannot go on-shell, suppressing the cross section. For our setup this occurs at roughly 400 GeV.
This simple kinematic argument however, has big implications for experimental searches in the full theory. One should endeavour to not cut away the region in which the signal peaks, therefore adjusting the minimum $\met$ cut to become a function of $M_\phi$ should provide a natural way to optimize signal over background.  

We now make some operator specific statements, firstly we note that the limits obtained on $\Lambda$ for the axial coupling are of the order $\mathcal{O}(1)$ TeV, and therefore vindicate the use of the EFT in these searches, however the results on $\Lambda$ are also bounded by below, since the EFT cannot in general be trusted for $\Lambda < 200$ GeV. Secondly we note that the limits on $\Lambda$ from the top-scalar operator are of the order $\Lambda \sim 150$ GeV \cite{Haisch:2012kf}. This is exactly in the region of $M_\phi\,,g\,,\Lambda$ phase space in which the EFT begins to breakdown, however, if the FT was used with slightly softer $\met$ cuts one obtains cross sections in which the EFT and FT are similar. It is clear that when setting limits on the top-scalar operator, the FT effects should be investigated, particularly since in this region the validity of the EFT is correlated with the exact phase space cuts applied.  Note, however, that over a large range of mediator mass the EFT actually underestimates the size of the DM production cross section, leading to the present LHC constraints, carried out in the EFT, being conservative.
We note that the limits set on $\Lambda$ by the charm- and bottom-scalar operators are in the region where the EFT is extremely dubious, since the cross section here is suppressed by around five orders of magnitude.  

\begin{figure}
\includegraphics[width=8cm]{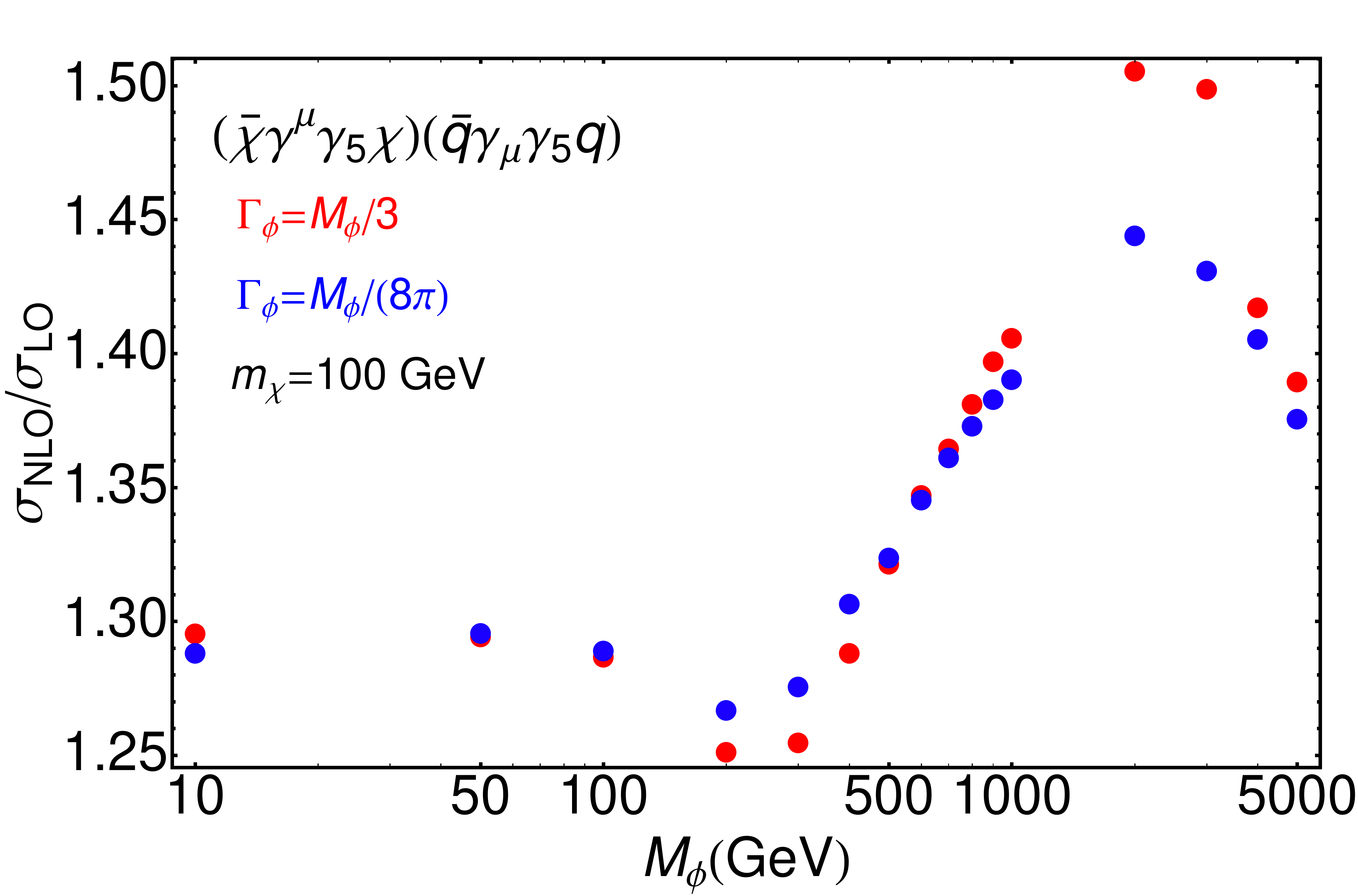}
\caption{The ratio of NLO to LO cross sections for the axial operator in the FT, for DM mass of 100 GeV. The red curve indicates the results 
obtained with width set to $M_{\phi}/3$ and the blue curve indicates the results obtained with a width of $M_{\phi}/(8\pi)$.}
\label{fig:FT_K}
\end{figure}

Although thus far we have only considered fixed widths, $\Gamma_\phi$, for very wide mediators a running width, $\tilde{\Gamma}(s)$, may be more appropriate~\cite{An:2012va}.  This corresponds to the replacement in the mediator propagator of,
\be
(s_{\bar{\chi}\chi}-M_\phi^2)+iM_\phi\Gamma_\phi \rightarrow (s_{\bar{\chi}\chi}-M_\phi^2)+i\sqrt{s_{\bar{\chi}\chi}}\,\tilde{\Gamma}(s_{\bar{\chi}\chi})~.
\ee
When the mediator is heavier than all the particles it couples to, the running width may be approximated by, $\tilde{\Gamma}(s) = \sqrt{s} \Gamma_\phi/M_\phi$.
Given the flexibility of our code it is straightforward to implement the above changes to include running width effects.  As an example, for the top scalar coupling and a mediator mass of 1 TeV we find the difference in the total cross section between the fixed and running width is sub-percent for the narrower case.  For the wide mediator, the effect is larger, as expected. The running width increases the cross section by $\sim 15\%$.

Finally, it is interesting to consider the impact of our NLO corrections on the axial operator in the full theory. Therefore we plot the ratio of NLO to LO cross sections as a function of $M_{\phi}$ for our two width choices in Fig.~\ref{fig:FT_K}. There is a clear dependence on $M_{\phi}$, with the $K$-factor growing as a function of $M_{\phi}$. This dependence is controlled by the width of the mediator and the resulting shape of the Breit-Wigner. As expected for heavy mediators the effective theory inclusive $K$-factor is obtained.

\section{Monophoton}\label{sec:monogamma}

In this section we describe the phenomenology associated with DM production in association with a monophoton. 
In order to define photons in a hadronic environment an isolation criteria must be applied. 
This isolation reduces the contributions from the production of secondary photons in the decays of certain types of hadrons. 
Typically an experimental isolation criterion requires the amount 
of hadronic energy inside a cone around the photon to be less than a fixed input, \ie
\begin{eqnarray}
\sum_{\mathrm{had}\in R_0} E_T^{\mathrm{had}} < E_T^{\mathrm{max}} \quad {\rm{with}} \quad R_0=\sqrt{\Delta\phi^2+\Delta\eta^2}\,.
\end{eqnarray}
We use typical values at the LHC for the isolation cuts, \ie\ $R_0= 0.5$ and $E_T^{\mathrm{max}}= 5\ \gev$.
On the theoretical side this form of isolation introduces complications at NLO since photons radiated, through bremsstrahlung, from final state
fermions induce a collinear singularity. This singularity has no corresponding 
singularity on the virtual side. In order to render the calculation finite the contributions from the fragmentation 
functions must be included. These fragmentation functions require non-perturbative input needed to provide boundary information at a certain scale. Our treatment of the fragmentation contributions is the same as in previous photon processes implemented in MCFM~\cite{Campbell:2011bn,Campbell:2012ft}, and we use the fragmentation functions of~\cite{Bourhis:1997yu}.  The vector and axial-vector production of DM along with a photon have been recently considered~\cite{Huang:2012hs}, although the effects of photon fragmentation were not included. 
\begin{figure}
\includegraphics[width=8cm]{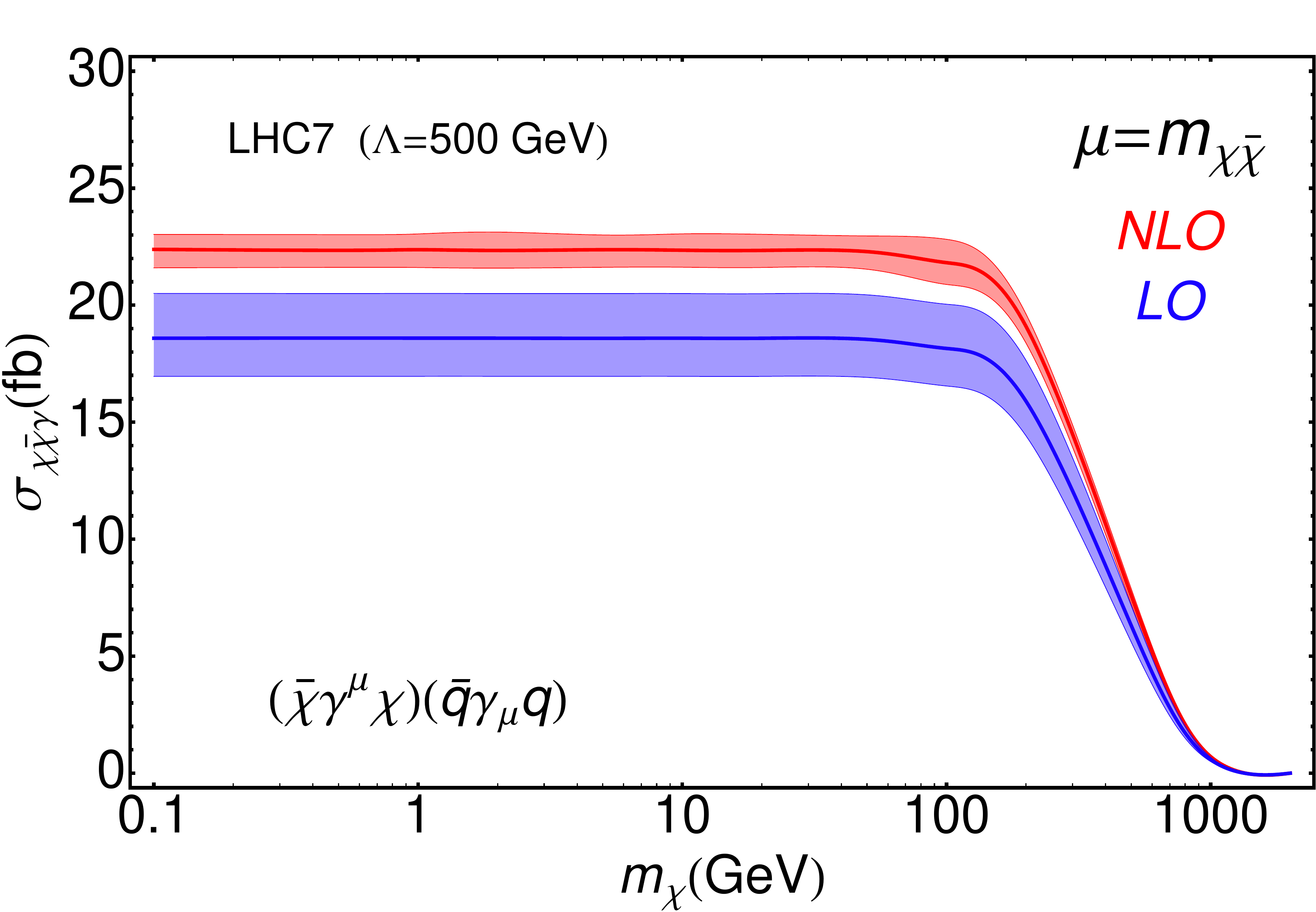}
\includegraphics[width=8cm]{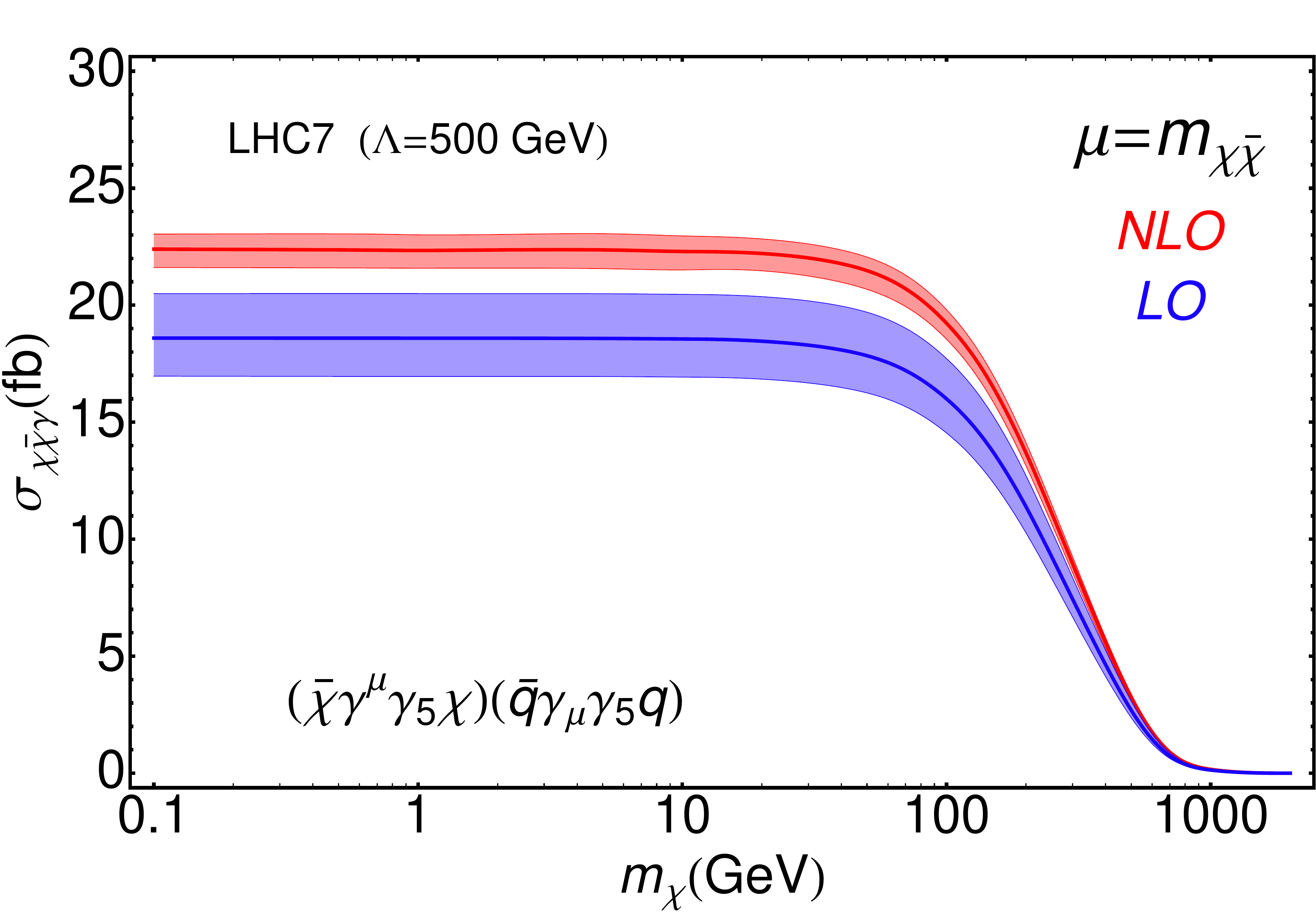}
\begin{center} 
\end{center}
\includegraphics[width=8cm]{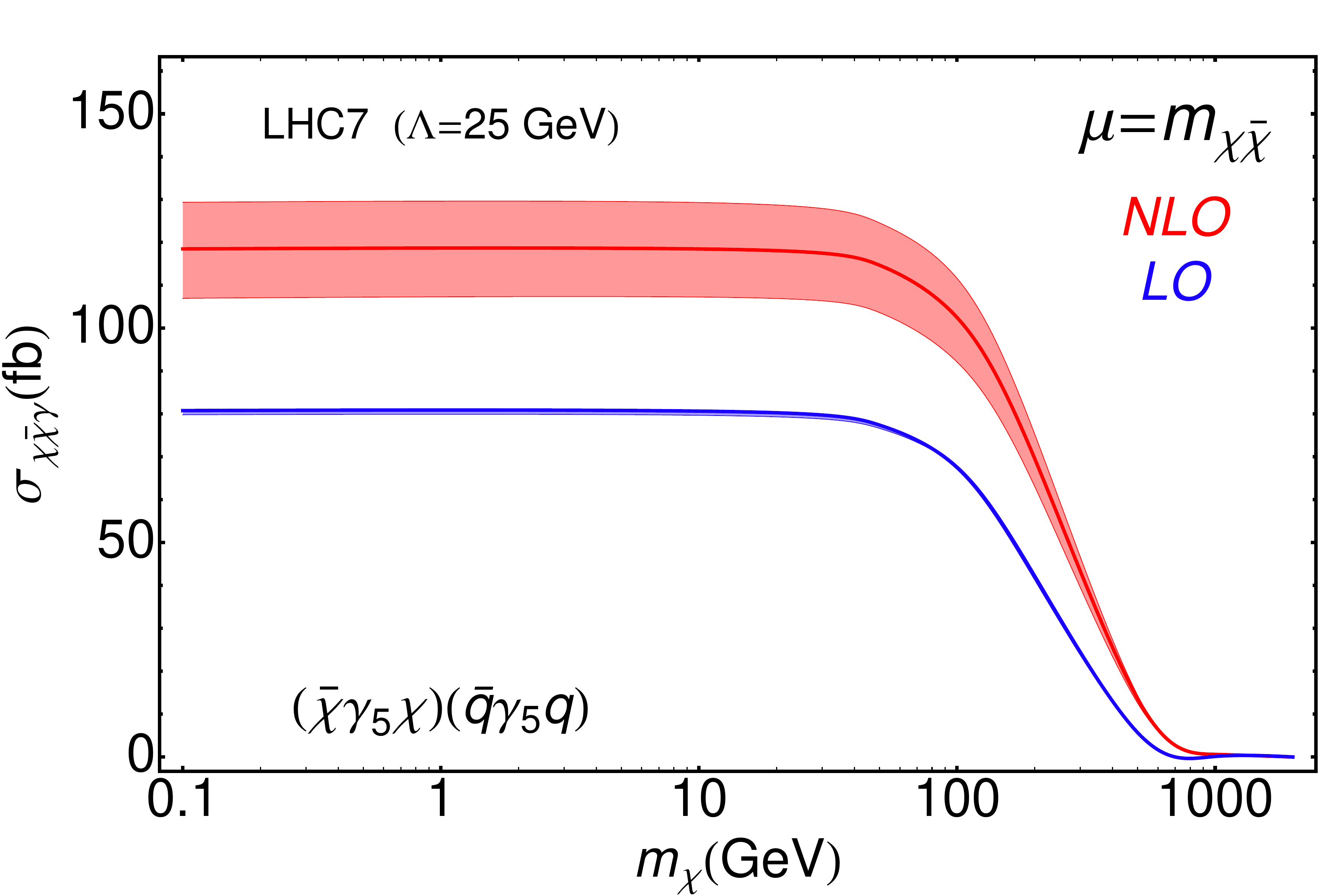}
\includegraphics[width=8cm]{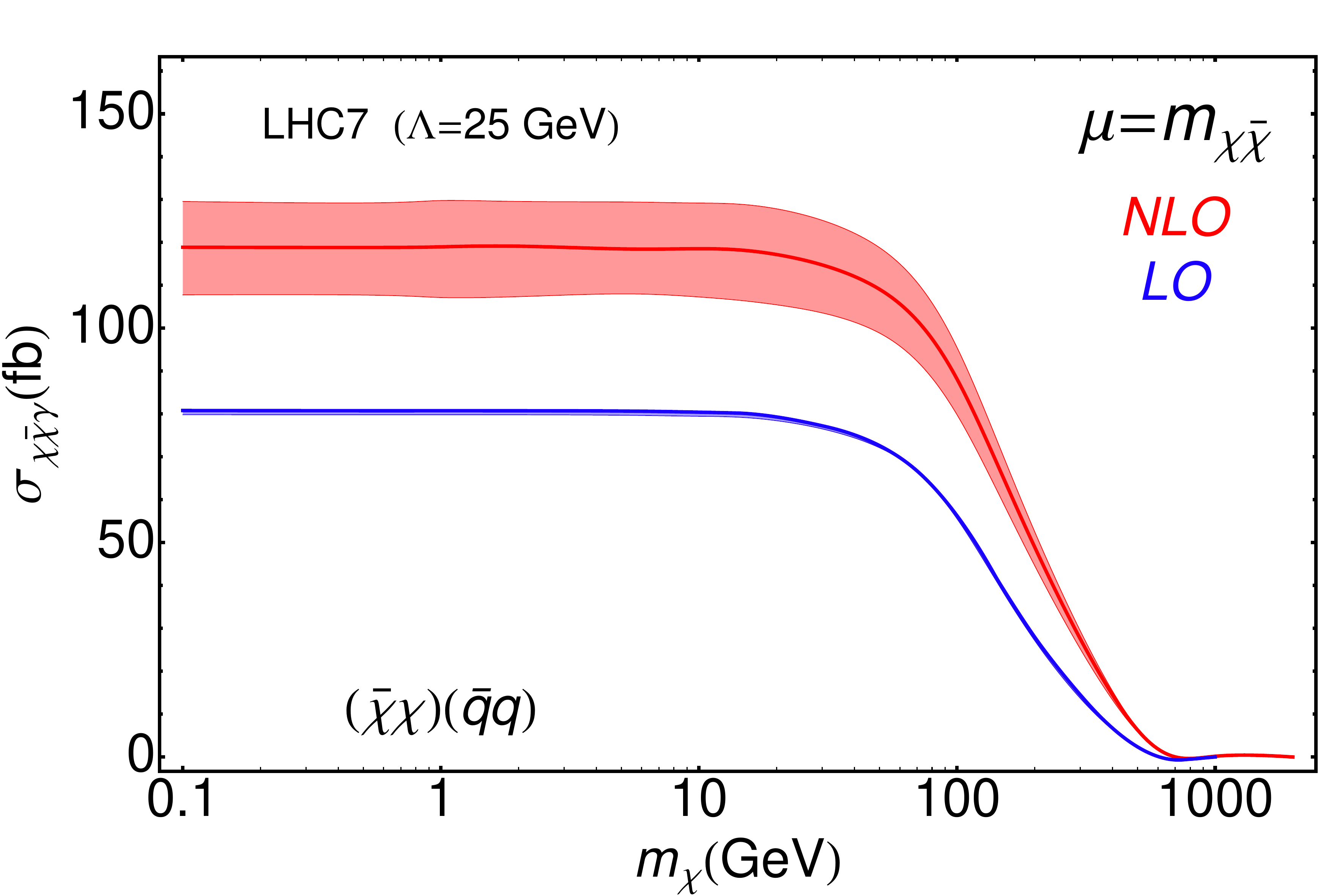}
\begin{center} 
\end{center}
\caption{LO and NLO cross sections for DM production in association with a single photon at the 7 TeV LHC. The solid line indicates the cross section obtained with the default scale $\mu=m_{\chi\overline{\chi}}$, the shaded band represents the deviation from this scale when the scales are varied by a factor of two in each 
direction. The phase space cuts described in the text (\ref{eq:monophotoncuts}) have been applied. }
\label{fig:xs_monophot}
\end{figure}

\subsection{Phenomenology} 

The procedure we follow for the monophoton examples we consider is very similar to that described earlier, section~\ref{sec:monojet}.  We base our cuts around those 
used in current experimental analyses~\cite{Chatrchyan:2012tea,Aad:2012fw}, more specifically we require our final state to satisfy, 
\begin{eqnarray} 
\label{eq:monophotoncuts}
\slashed{E}_T > 140  \,\,{\rm{GeV}}\,,\; \quad p^{\gamma}_{T} > 150  \,\,{\rm{GeV}}\,, \; \quad |\eta_\gamma| < 2\,.
\end{eqnarray}

\begin{figure}
\includegraphics[width=8cm]{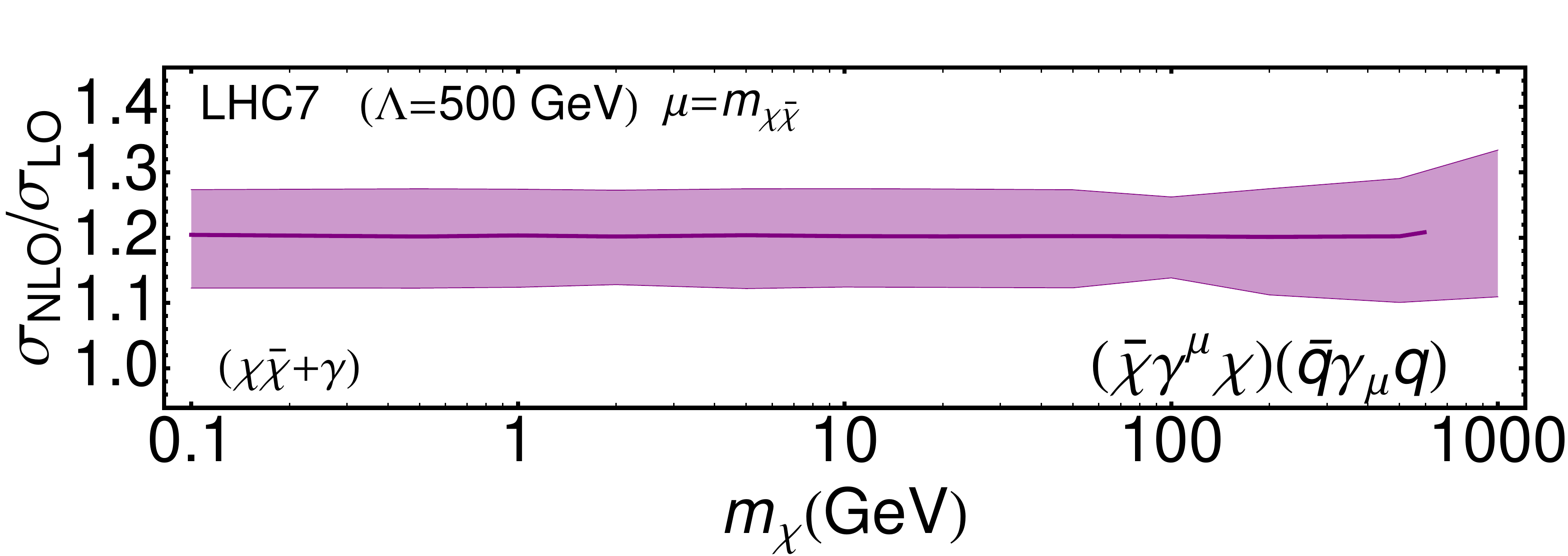}
\includegraphics[width=8cm]{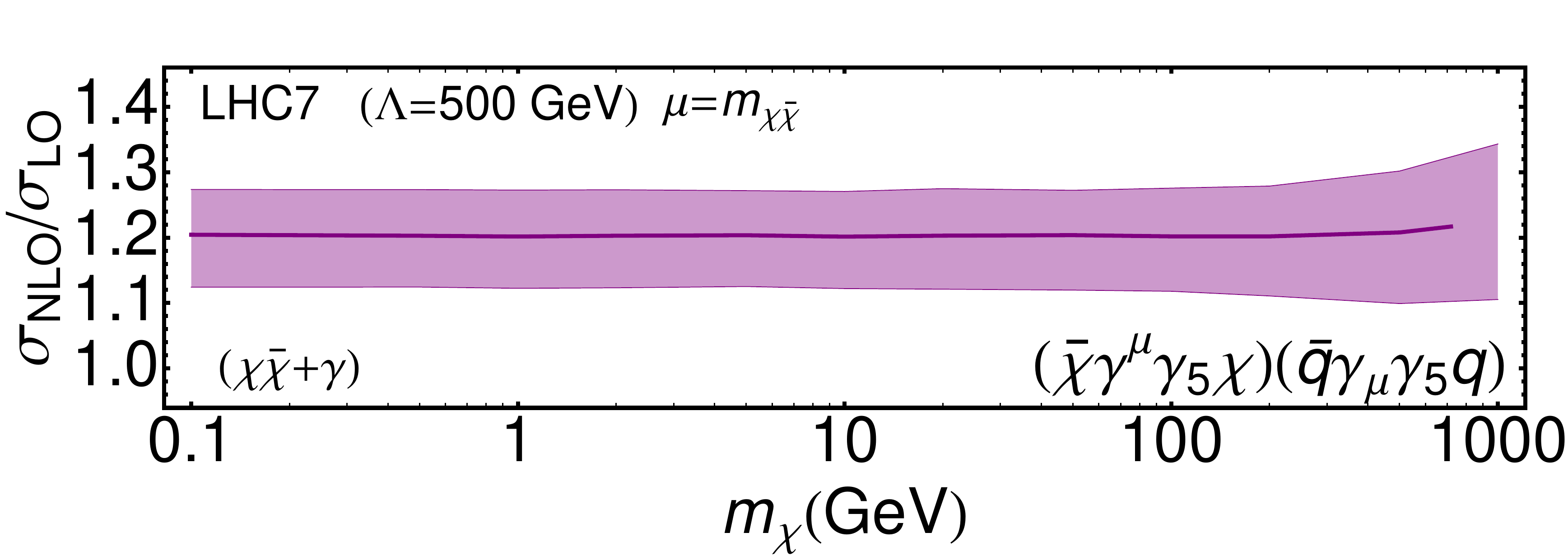}
\begin{center} 
\end{center}
\includegraphics[width=8cm]{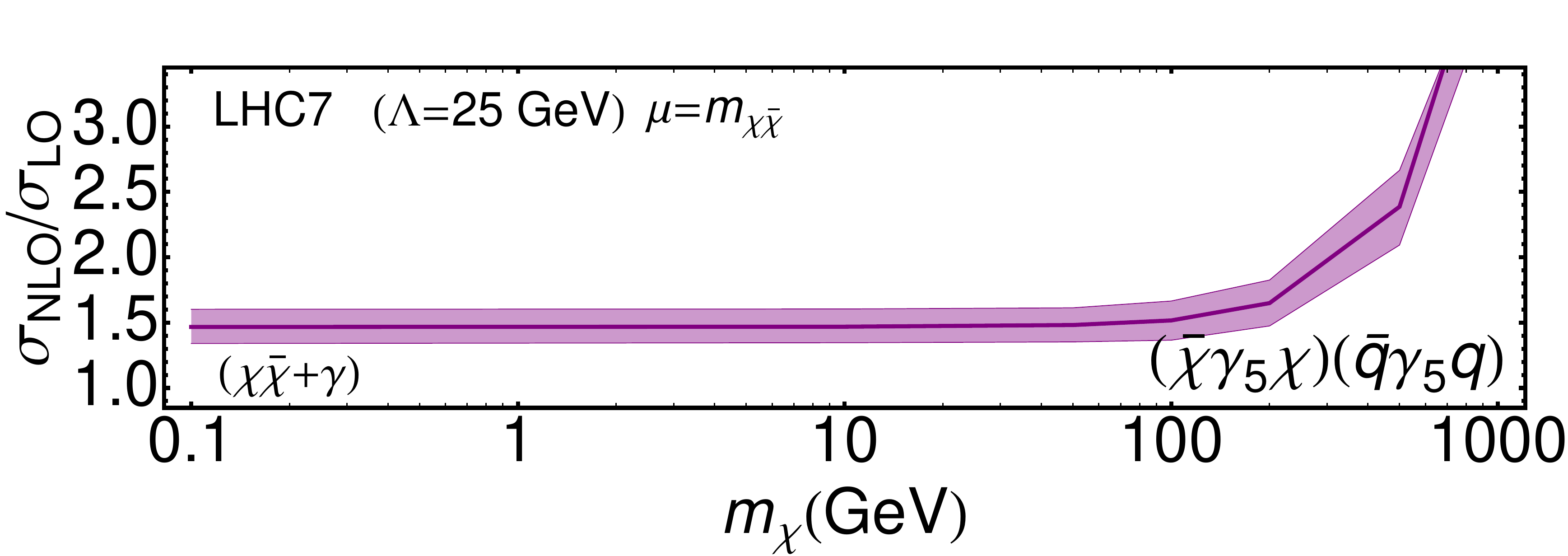}
\includegraphics[width=8cm]{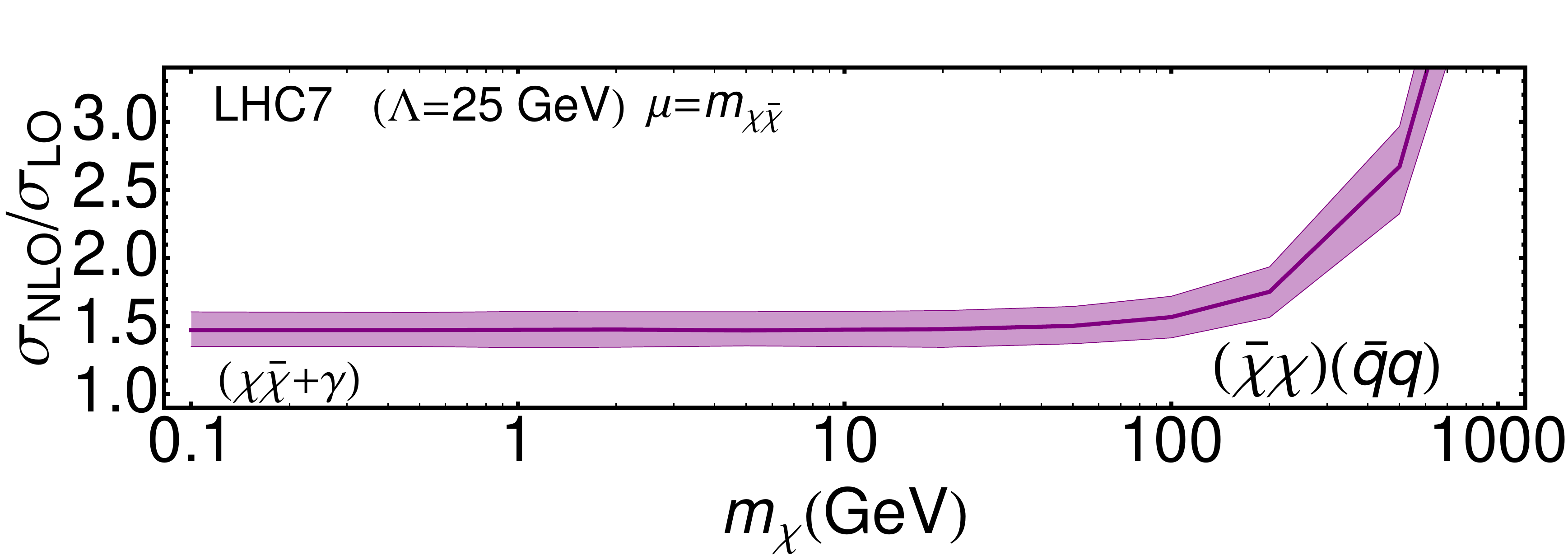}
\begin{center} 
\end{center}
\caption{$K$-factors for DM production in association with a photon at the 7 TeV LHC. The solid line indicates the $K$-factor obtained with the default scale $\mu=m_{\chi\overline{\chi}}$, the shaded band represents the $K$-factors obtained using scales varied by a factor of two in each direction. The phase space cuts described in the text have been applied. }
\label{fig:K_monogam}
\end{figure}

We present LO and NLO cross sections, under these cuts for various operators in Fig.~\ref{fig:xs_monophot}, $K$-factors obtained for these operators are illustrated in Fig.~\ref{fig:K_monogam}. We observe that, as was the case for monojet examples, the axial and vector results are similar to each other, and the scalar and pseudo-scalar are likewise similar.  The axial and vector examples have an $K$-factor of around 1.2, which corresponds to an increase in $\Lambda$ of around 5\%. The scale variation for these operators displays a similar behaviour to the background $Z\gamma$ process~\cite{Campbell:2011bn}. These processes have no $\alpha_S$ dependence at LO, therefore the only scale dependence enters through the factorization scale dependence associated with the PDFs. As a result, the cross section increases with increasing $\mu$ which is the opposite dependence to the renormalization scale which decreases the cross section as $\mu$ increases. At NLO one may have expected a large scale dependence for this process, since it is LO in the renormalization scale.  However, there is a net cancellation between variations in the factorization scale and renormalization scales such that the scale dependence for $V\gamma$ at NLO is very small. The axial and vector operators inherit this trait, at LO the (factorization) scale dependence is around $9\%$ whilst at NLO the scale variation is around $2-3\%$. As is clear from the plots, the variation around the default scale at LO does not include the NLO curve so some care should be taken in interpreting this small variation as a small theoretical uncertainty. 

\begin{figure}
\includegraphics[width=8cm]{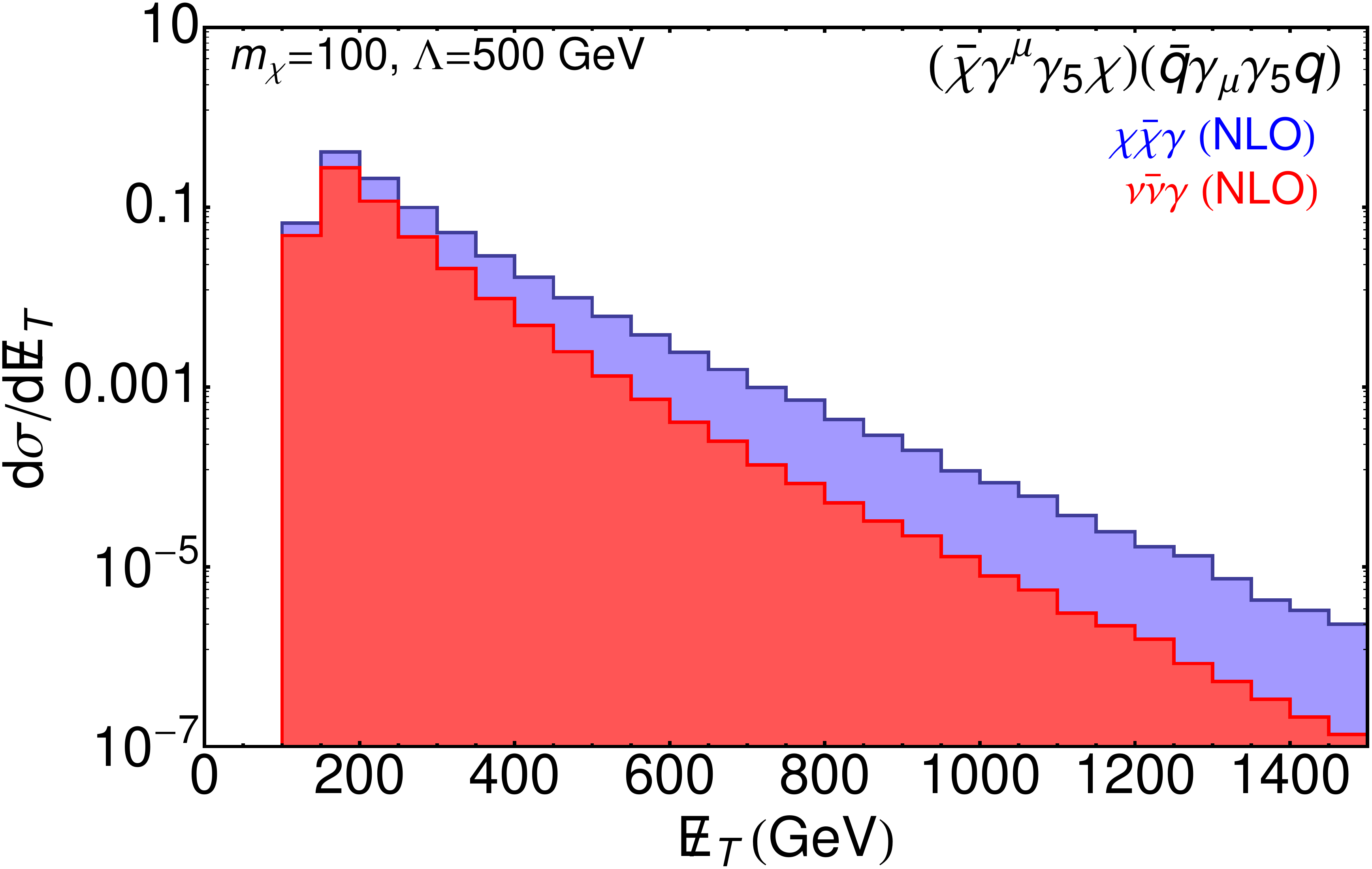}
\caption{NLO predictions for the missing transverse momentum spectrum for signal and background obtained using MCFM, for the monophoton process, proceeding via an axial operator. The DM mass is 100 GeV, and the scales have been chosen to be equal to the DM invariant mass $\mu=m_{\chi\overline{\chi}}$.}
\label{fig:ET_monogam}
\end{figure}

It is immediately obvious that the scalar and pseudo-scalar operators do not share the same traits as the other operators when it comes to scale variation. This is also naturally explained from the discussion in the previous paragraph. Here the heavy sea quarks have a very small factorization scale dependence, and as such the LO result barely depends upon the scale choice ($< 1$\%). However, at NLO the lack of a factorization scale dependence in a significant part of the calculation removes the accidental cancellation. As a result the renormalization scale dependence dominates and there is a larger scale dependence at NLO around $(10\%)$. As was the case for the monojet study the $K$-factor for these operators is larger (1.5-2), however one expects large NNLO corrections for these operators since the contributions from two-glue initial states should be comparatively large. Note that the $\Lambda$ limit for these operators is expected to be very small, due to the quark mass suppression, and one expects the full theory to suppress this even further. 

\begin{figure}
\includegraphics[width=8cm]{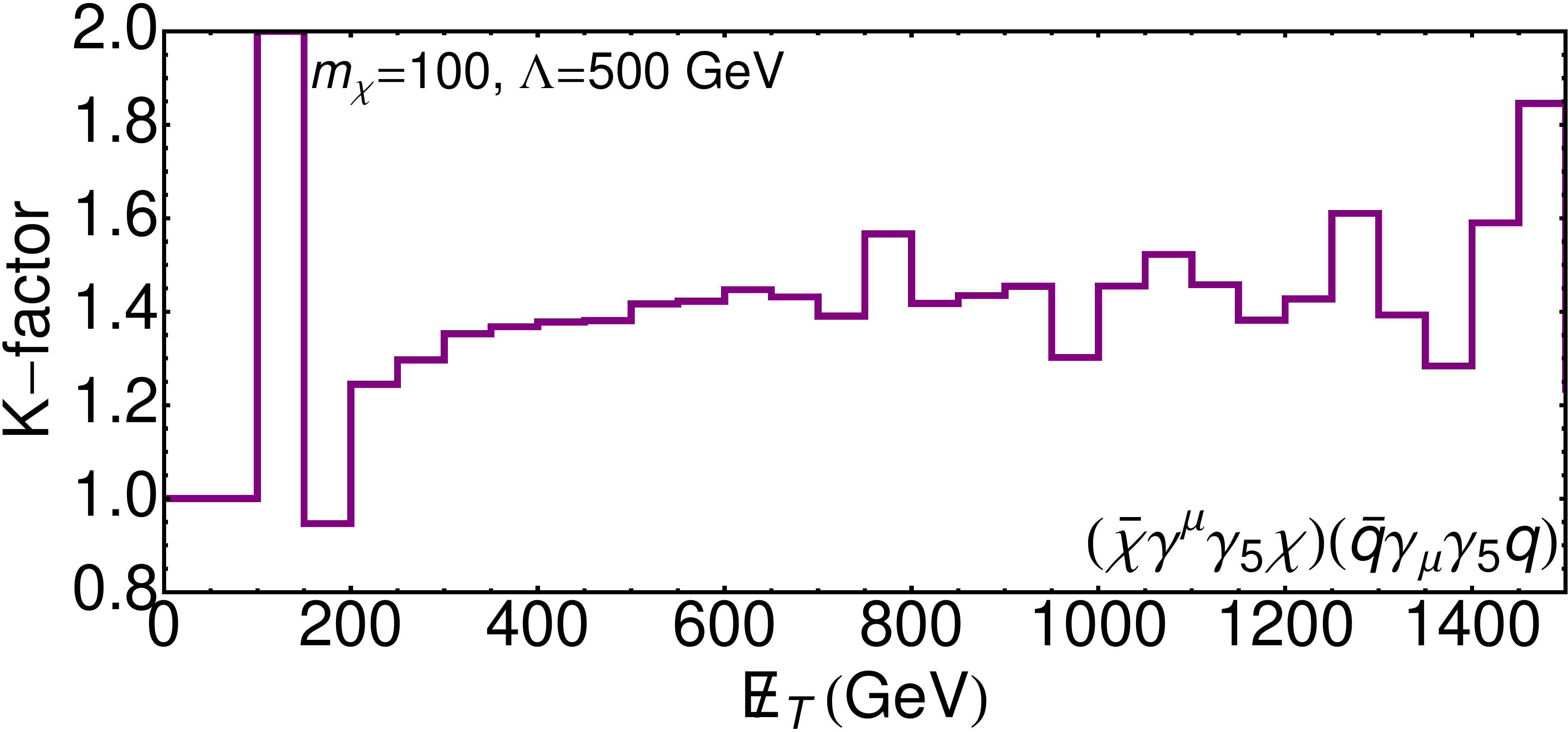}
\caption{The differential $K$-factor for monophoton production via an axial operator for $m_{\chi}=100$ GeV.}
\label{fig:diffK_monogam}
\end{figure}

We consider the differential $\met$ spectrum at NLO, as an example we consider the axial operator with $m_{\chi}=100 $ GeV. 
Again we compare to the shape of the dominant background (in this case $Z\gamma$) using MCFM to obtain both spectra at parton level. Our results for the spectrum and LO to NLO ratios are shown in Fig.~\ref{fig:ET_monogam} and Fig.~\ref{fig:diffK_monogam}. Since our explanation of the hardening of the spectrum for the signal in the monojet case did not invoke any properties of the recoil object we expect the signal to also be harder for this operator. This is indeed what we observe, the spectrum for the DM signal is significantly harder than that of the $Z\gamma$ background.  Away from the first bin the $K$-factor is also fairly stable as a function of the $\met$.
The mismatch of the photon $p_T$ and the $\met$ cuts means that at LO there is no contribution to the first bin, since $p_T$ balance enforces that the larger photon $p_T$ cut is also applied to the $\met$.  However, at NLO there can be events, with lower $\met$, in the first bin.  This accounts for the large (actually infinite) K-factor in the first bin. At NLO the full fiducial phase space is explored, ($\slashed{E}_T >$ 140) resulting in a non-zero cross section in the first $\slashed{E}_T$ bin. 

Finally we note that 
care should be taken when estimating scale variations for these processes if a jet-veto is applied. Additional information on the topic of scale variation under the application of jet-vetos can be found in Refs.~\cite{Stewart:2011cf,Campbell:2012ft}.

\section{Conclusions} 
\label{sec:conclusions}

Hadron colliders provide an ideal place to search for DM, offering complementary results to those obtained from direct 
detection experiments.  Model independent searches for the pair production of DM require some other visible activity in the event, 
\eg\ jets, photons, or vector bosons.  We focussed on the cases of monojet and monophoton, for which there are existing searches both at the Tevatron and the LHC.  Colliders are free from astrophysical uncertainties and are competitive with the direct detection experiments for light dark matter and DM with spin-dependent nucleus couplings.  One of the dominant sources of uncertainty of the experimental results from the LHC is systematic uncertainties.  This is already true for the monojet searches from the 2011 data set, and one 
assumes that this will continue for 2012 data.  In this regard, it is important for the theoretical community to aid the experimental effort in any way it can 
in reducing these uncertainties. We addressed one such systematic error, the uncertainty associated with the precise rate for the signal process. 
Leading Order predictions suffer from a large dependence on the unphysical renormalization and factorization scales. 
Calculating at NLO accuracy curbs this bad behaviour somewhat, with the additional happy byproduct that, since the NLO 
corrections typically introduce a $K$-factor $>$ 1, limits using  the NLO prediction will be stronger in addition to having smaller 
systematic uncertainties.  

To remain as general as possible, and to allow extension of our results to other similar processes, we factorized the SM production from the BSM decays.  We provided analytic expressions both for the SM production at NLO and the LO decay amplitudes 
of the DM particles.  The exact nature of the SM production (and DM decay) depends on the particle mediating the interactions, and we presented results for interactions which were mediated by vector, axial, scalar or pseudo-scalar particles.  We considered both the case where the mediating particle was very heavy and the effective coupling between DM and the SM was a contact operator and the case of a propagating mediator. We implemented all our results into MCFM and they will be made available in the next release\footnote{Advanced copies can be obtained from the authors, upon request}.  This implementation is sufficiently versatile that it can handle all the operators considered here, as well as more general (parity violating) variations, as well as more general UV completions

We used our NLO results to study the monojet and monophoton phenomenology in the presence of a DM signal. We presented 
NLO and LO cross sections and demonstrated there is an increase in rate as well as a reduction in the renormalization/factorization scale dependence.  Thus reducing the systematic uncertainty in the signal prediction.  Using similar cuts to the experimental analyses we were able to infer the amount of improvement on the limits on $\Lambda$ obtained by using a NLO prediction for the total rate. Typically these limits were increased by around $5-20\%$ with the exact results being highly operator dependent.  Using these NLO results we were able to confirm earlier results that saw a significant hardening of the signal $\met$ spectrum, compared to the dominant irreducible background.  Furthermore, we found a slight enhancement of this feature at NLO due to a rising differential $K$-factor.

We also investigated some of the effects of re-instating the mediating particle for the monojet case. This confirmed the validity of the EFT for the axial theory. However we saw that the region of $\Lambda$ probed at the LHC by the top-scalar operator is in a region in which the differences between the full theory and the EFT are extremely sensitive to the mediator mass and the exact fiducial phase space cuts. As a result we suggest that, should the collaborations proceed to set constraints on this operator, they investigate the typical differences between the FT and EFT in order to ensure the validity of the EFT results. 

Going forward, as the LHC collects more data at both higher energies and higher instantaneous luminosity, it will continue to constrain the dark sector.  As running conditions change so must analyses and having signal predictions at NLO, through MCFM, will allow the analysis cuts to be tuned to best separate signal events from background and maximize the potential for finding DM at the LHC.

\section*{Acknowledgements} 

We thank John Campbell, Keith Ellis and Roni Harnik for useful discussions. Fermilab is operated by Fermi Research Alliance, LLC under
Contract No. DE-AC02-07CH11359 with the United States Department of Energy.

\appendix
\section{Spinor Helicity Formalism}
\label{app:spin} 
In this appendix we define our spinor products, 
(for a review see ref.~\cite{Dixon:1996wi})
The function $u_\pm(k_i)$ represents a
massless Weyl spinor of momentum $k_i$ and positive or negative
chirality. In terms of these solutions to the Dirac equation, the spinor products are 
defined by,
\begin{eqnarray}
\spa i.j &=& 
              \langle i^-|j^+\rangle = \bar{u}_-(k_i)u_+(k_j)\,,
             \label{defspa}\\
\spb i.j &=& 
             \langle i^+|j^-\rangle = \bar{u}_+(k_i) u_-(k_j)\,.
             \label{defspb}
\eea
We use the convention $\spb{i}.{j} = \mathop{\rm
sgn}(k_i^0k_j^0)\spa{j}.{i}^*$, so that,
\be
\spa{i}.{j}\spb{j}.{i} = 2k_i\cdot k_j \equiv s_{ij} \,.
\ee
\be
\langle a|i|b]\ =\ \spa{a}.{i}\spb{i}.{b} \,, \qquad\quad
\langle a|(i+j)|b]
\ =\ \spa{a}.{i}\spb{i}.{b} + \spa{a}.{j}\spb{j}.{b} \,.
\label{spinorsandwich}
\ee
Further useful identities are,
\be
\langle i^{\pm}|\gamma^\mu |i^\pm\rangle = 2k_i^\mu\,, \quad \langle i | \gamma^\mu| j]\langle k | \gamma_\mu| l] = 2\spa{i}.{k}\spb{l}.{j}~.
\ee
We will also need the following one-loop basis functions, those associated with the reduction of tensor triangles,
\begin{eqnarray}
\rm{L}_0(x,y)=\frac{\ln{(x/y)}}{1-\frac{x}{y}}, \quad
\rm{L}_1(x,y)=\frac{\rm{L}_0(x,y)+1}{1-\frac{x}{y}} 
\end{eqnarray}
and the finite part of the one-mass box given by 
\begin{eqnarray}
\rm{Ls}_{-1}\bigg(\frac{x_1}{y_1},\frac{x_2}{y_2}\bigg)=\Li_{2}\bigg(1-\frac{x_1}{y_1}\bigg)+\Li_{2}\bigg(1-\frac{x_2}{y_2}\bigg)+\ln{\bigg(\frac{x_1}{y_1}\bigg)}\ln{\bigg(\frac{x_2}{y_2}\bigg)}-\frac{\pi^2}{6}
\end{eqnarray}

\section{NLO Vector Currents}
\label{app:VecC}
We follow the notation of \cite{Bern:1997sc} and decompose the amplitude as follows, 
\begin{eqnarray}
\mathcal{A}^{(1,\mu)}_V=c_{\Gamma} ( \mathcal{A}^{(0,\mu)}_V V + F^{\mu})~,
\end{eqnarray}
where the ubiquitous one-loop prefactor is defined as 
\begin{eqnarray}
c_{\Gamma}=\frac{1}{(4\pi)^{2-\epsilon}}\frac{\Gamma(1-\epsilon)^2\Gamma(1+\epsilon)}{\Gamma(1+2\epsilon)}~.
\end{eqnarray}
The amplitudes we will write down presently have not been UV-renormalised, in order to obtain UV finite results one must perform 
a UV-subtraction, e.g. in the $\overline{\rm{MS}}$-scheme one should subtract, 
\begin{eqnarray}
c_{\Gamma} N_c g^2 \bigg(\frac{1}{\epsilon}\bigg(\frac{11}{3}-\frac{2}{3}\frac{n_f}{N_c}\bigg)\bigg) \mathcal{A}^{(0,\mu)}_{V}.
\end{eqnarray}
We are now in a position to write down the virtual corrections to the monojet process, the leading colour contributions have the following form, 
\begin{eqnarray}
\mathcal{A}^{1lc,\mu}_{V}(1^+_q,2^+_g,3^-_{\overline{q}})=c_{\Gamma} ( \mathcal{A}^{(0,\mu)}_V V^{lc} + F^{lc,\mu})~,
\end{eqnarray}
with 
\begin{eqnarray}
V^{lc}=-\frac{1}{\epsilon^2}\bigg(\bigg(\frac{\mu^2}{-s_{12}}\bigg)^{\epsilon} +\bigg(\frac{\mu^2}{-s_{23}}\bigg)^{\epsilon}\bigg)-\frac{3}{2\epsilon}\bigg(\frac{\mu^2}{-s_{23}}\bigg)^{\epsilon} - 3, 
\end{eqnarray}
and 
\begin{eqnarray}
F^{lc,\mu}(1_{q}^+,2^+_g,3^{-}_{\overline{q}})=-\mathcal{A}^{(0,\mu)}_V  {\rm{Ls}}_{-1}\bigg(\frac{-s_{12}}{-s_{123}},\frac{-s_{23}}{-s_{123}}\bigg)+\frac{1}{2}\frac{\spaa 3.\gamma^{\mu}.1.3}{\spa1.2\spa2.3}{\rm{L}}_0(s_{23}/s_{123})\nonumber\\+\frac{1}{4}\frac{\spa1.3^2\spbb 1.\gamma^{\mu}.(2+3).1}{\spa1.2\spa2.3}\frac{{\rm{L}}_1(s_{23}/s_{123})}{s_{123}}~.
\label{eq:fVlc}
\end{eqnarray}
The subleading in colour amplitude has the following form, 
\begin{eqnarray}
V^{sl}=-\frac{1}{\epsilon^2}\bigg(\frac{\mu^2}{-s_{12}}\bigg)^{\epsilon} -\frac{3}{2\epsilon}\bigg(\frac{\mu^2}{-s_{123}}\bigg)^{\epsilon} - \frac{7}{2}~.
\end{eqnarray}
The remaining $F^{sl,\mu}$ pieces are, 
\begin{eqnarray} 
&&F^{sl,\mu}(1_{q}^+,2^+_g,3^{-}_{\overline{q}}) = \frac{\spaa 3.(1 + 2).\gamma^{\mu}.3}{2 \spa1.2\spa 2.3} {\rm{Ls}}_{-1}\bigg(\frac{-s_{12}}{-s_{123}},\frac{-s_{13}}{-s_{123}}\bigg)
\nonumber\\&&+\frac{1}{2}\frac{\spa1. 3^2 \spaa2. (1 + 3). \gamma^{\mu}. 2}{
  \spa1. 2^3 \spa2. 3} {\rm{Ls}}_{-1}\bigg(\frac{-s_{13}}{-s_{123}},\frac{-s_{23}}{-s_{123}}\bigg)  -\frac{ \spb1. 2 \spaa3. (1 + 2). \gamma^{\mu}. 1}{
   \spa1. 2 s_{123}}{\rm{L}}_0(s_{23},s_{123})
   \nonumber\\
   &&- \frac{1}{4}\frac{\spaa1. (2 + 3). \gamma^{\mu}. 1 \spb1. 2^2 \spa2. 3}{
  \spa1. 2}\frac{{\rm{L}}_1(s_{123},s_{23})}{s_{23}^2} + \frac{1}{2}\frac{\spaa1. (2 + 3). \gamma^{\mu}. 1 \spa2. 3 \spb2. 1}{
  \spa1. 2^2}\frac{{\rm{L}}_0(s_{123},s_{23})}{s_{23}} \nonumber\\
 && -\frac{1}{2} \frac{\spa3. 1 \spb1. 2 \spab2. \gamma^{\mu}. 2 s_{123}}{
  \spa1. 2} \frac{{\rm{L}}_1(s_{123},s_{13})}{s_{13}^2} + 
\frac{1}{2}\frac{ \spa3. 1 \spb1. 2 \spaa2.(1 + 3). \gamma^{\mu}. 1}{
  \spa1. 2^2} \frac{{\rm{L}}_0(s_{123},s_{13})}{s_{13}} \nonumber\\
&&   + \frac{
 \spb1. 2 \spbb3. (1 + 2). \gamma^{\mu}. 2 - 
  \spb2. 3 \spbb1.(3 + 2). \gamma^{\mu}. 2}{
 4 \spb1. 3 \spb2. 3 \spa1. 2}~.
\end{eqnarray}
Finally, we note that for the vector current the diagrams associated with closed fermion loops vanish via Furrys theorem, when constructing axial currents  
we will need the following non-vanishing $n_F$ axial current
\begin{eqnarray}
F^{ax,\mu}(1_{q}^+,2^+,3^{-}_{\overline{q}}) =-\frac{1}{2}\frac{\spab3.\gamma^{\mu}.2 \spb2.1 {\rm{L}}_1(s_{13},s_{123})}{s_{123}}~.
\end{eqnarray}
We note that the above results can be checked by contraction with the current in $\spab4.\gamma_{\mu}.5/s_{45}$, reproducing the formulae listed in \cite{Bern:1997sc}. 

At NLO we also require the tree-level amplitudes involving the emission of an additional parton.  The necessary (with the remaining helicity assignments being obtained via line-reversal and conjugation) two gluon amplitudes are, 
\begin{eqnarray}
\mathcal{A}^{(0,\mu)}_{V}(1^+_q,2^+_g,3^+_g,4^-_{\overline{q}})& =& -\frac{\spaa4.(1+2+3).\gamma^{\mu}.4}{2\spa1.2\spa2.3\spa3.4}~, \\
\mathcal{A}^{(0,\mu)}_{V}(1^+_q,2^+_g,3^-_g,4^-_{\overline{q}})&=&-\frac{\spa3. 1 \spb1. 2 \spaa3. (1 + 2). \gamma^{\mu}. 4}{
  2 \spa1. 2 s_{23} s_{123}} \nonumber\\
&&+ \frac{
 \spa3. 4 \spb4. 2 \spbb2. (3 + 4). \gamma^{\mu}. 1}{
 2 \spb3. 4 s_{23} s_{234}} 
- 
 \frac{\spab3|(1 + 2).\gamma^{\mu}.{(3 + 4)|2}}{
 2 \spa1. 2 \spb3. 4 s_{23} }~,	 \\
 \mathcal{A}^{(0,\mu)}_{V}(1^+_q,2^-_g,3^+_g,4^-_{\overline{q}})&=&
\frac{\spb1. 3^2 \spaa2.(1 + 3). \gamma^{\mu}. 4}{
  2 \spb1. 2 s_{23} s_{123}} \nonumber\\&&- \frac{
  \spa2. 4^2 \spbb3. (2 + 4). \gamma^{\mu}. 1}{
  2 \spa3. 4 s_{23} s_{234}}  - \frac{
  \spb1. 3 \spa2. 4 \spab4. \gamma^{\mu}. 1}{
  2 \spb1. 2 \spa3. 4 s_{23}}~.
\end{eqnarray}
The four-quark amplitudes are 
\begin{eqnarray}
\mathcal{A}^{(0,\mu)}_{V}(1^+_q,2^+_{\overline{Q}},3^-_{Q},4^-_{\overline{q}})&=&-\frac{1}{2} \bigg(\frac{\spb 1.2 \spaa 3.(1 + 2). \gamma^{\mu}. 4}{
    s_{123}s_{23}} + 
   \frac{\spa3.4 \spbb2.(3 + 4).\gamma^{\mu}.1}{s_{234} s_{23}}\bigg)~, \\
   \mathcal{A}^{(0,\mu)}_{V}(1^+_q,2^-_{\overline{q}},3^-_{Q},4^+_{\overline{Q}})&=&\mathcal{A}^{(0,\mu)}_{V}(1^+_q,2^-_{\overline{q}},4^+_{Q},3^-_{\overline{Q}})~.
   \end{eqnarray}
For the monophoton calculation we will also need the following tree-level amplitudes, 
\begin{eqnarray}
\mathcal{A}^{(0,\mu)}_{V}(1^+_q,2^+_g,3^-_{\overline{q}},4^+_{\gamma})& =& -\frac{\spa1.3\spaa 3.(1+2+4).\gamma^{\mu}.3}{2\spa1.2\spa1.4\spa2.3\spa3.4}~, \\ 
\mathcal{A}^{(0,\mu)}_{V}(1^+_q,2^+_g,3^-_{\overline{q}},4^-_{\gamma})&=&\mathcal{A}^{(0,\mu)}_{V}(1^+_q,4^-_g,2^+_g,3^-_{\overline{q}})+\mathcal{A}^{(0,\mu)}_{V}(1^+_q,2^+_g,4^-_g,3^-_{\overline{q}})~.
\end{eqnarray}
In the above equations we have defined the photon as $p_4$ to emphasize that it is not colour-ordered.

\bibliography{DM_NLO}

\end{document}